\documentclass[preprint,12pt]{elsarticle}

\usepackage{amssymb}
\usepackage{amsthm}
\usepackage{nomencl}
\usepackage{amsmath}
\usepackage{algpseudocode}%
\usepackage[linesnumbered,ruled,vlined,noend]{algorithm2e}
\usepackage{xcolor}
\usepackage{float}
\usepackage[hidelinks]{hyperref}
\usepackage{hyperref}
\usepackage{url}

\newcommand{\mycomment}[1]{}

\journal{Journal}

\begin{document}

\begin{frontmatter}

\title{Quantum-Based Salp Swarm Algorithm Driven Design Optimization of Savonius Wind Turbine-Cylindrical Deflector System}

\author[1]{Paras Singh}
\author[1]{Vishal Jaiswal}
\author[1]{Subhrajit Roy}
\author[1]{Aryan Tyagi}
\author[1]{Gaurav Kumar*}
\author[1]{Raj Kumar Singh}

\affiliation[1]{organization={Department of Mechanical Engineering, Delhi Technological University},
            city={Delhi},
            postcode={110042}, 
            country={India}}

\cortext[cor1]{Corresponding author \\ \hfill E-mail address: \href{mailto:gauravkmr716@gmail.com}{gauravkmr716@gmail.com}}

\begin{abstract}
Savonius turbines, prominent in small-scale wind turbine applications operating under low-speed conditions, encounter limitations due to opposing torque on the returning blade, impeding high efficiency. A viable solution involves mitigating this retarding torque by directing incoming airflow through a cylindrical deflector. However, such flow control is highly contingent upon the location and size of the cylindrical deflector, and its angular velocity. This study introduces a novel design optimization framework tailored for enhancing the turbine-deflector system's performance. Leveraging surrogate models for computational efficiency, six different models were assessed, with Kriging selected for subsequent analysis based on its superior performance at approximating the relation between design parameters and objective function. The training data for the surrogate model and the flow field data around the system were obtained through Unsteady Reynolds-Averaged Navier Stokes (URANS) simulations using a sliding mesh technique. An in-house code for the Quantum-based Salp Swarm Optimization (QSSO) algorithm was then employed to obtain design parameters corresponding to the peak power coefficient ($C_p$) for the stationary deflector-turbine system. Additionally, the QSSO algorithm was quantitatively compared with nine other competing algorithms. The optimized stationary deflector-turbine system showed an improvement of $26.94\%$ in $C_p$ at Tip Speed Ratio (TSR) of $0.9$ compared to the baseline case. Further investigation into the effect of deflector rotational velocity ($\omega_d$) revealed significant improvements: $40.98\%$ and $11.33\%$ enhancement at $\omega_d$ = $3$ $rad/s$, and $51.23\%$ and $19.42\%$ at $\omega_d$ = $40$ $rad/s$, compared to configurations without a deflector and with the optimized stationary deflector, respectively at a TSR of $0.9$. In conclusion, this study introduces a robust optimization framework that not only improves the performance of Savonius turbines but also underscores the potential of surrogate modeling and advanced optimization algorithms in addressing aerodynamic challenges within wind turbine design. Importantly, the framework's applicability extends beyond this study, offering opportunities for multi-parameter optimization of components across the energy sector.

\end{abstract}

\begin{keyword}

Wind Turbines \sep Surrogate Modeling \sep Computational Fluid Dynamics \sep Metaheuristic Optimization \sep Swarm  Intelligence \sep Quantum Computing


\end{keyword}

\end{frontmatter}


\makenomenclature
\nomenclature{\(P\)}{Power available in incoming wind [$W$]}
\nomenclature{\(\rho\)}{Air density [$kg/m^3$]}
\nomenclature{\(V\)}{Instantaneous velocity [$m/s$]}
\nomenclature{\(V_\infty\)}{Free stream velocity [$m/s$]}
\nomenclature{\(A\)}{Area swept by turbine rotors [$m^2$]}
\nomenclature{\(\omega_t\)}{Angular velocity of turbine [$rad/s$]}
\nomenclature{\(\omega_d\)}{Angular velocity of deflector [$rad/s$]}
\nomenclature{\(\theta\)}{Azimuth Angle of turbine [$^o$]}
\nomenclature{\(R\)}{Radius of turbine rotor [$m$]}
\nomenclature{\(M\)}{Moment acting on turbine rotors [$Nm$]}
\nomenclature{\(C_p\)}{Power coefficient [-]}
\nomenclature{\(CoP\)}{Pressure coefficient [-]}
\nomenclature{\(C_m\)}{Moment coefficient [-]}
\nomenclature{\(\lambda\)}{Tip Speed Ratio (TSR) [-]}
\nomenclature{\(D_b\)}{Turbine blade diameter [$m$]}
\nomenclature{\(D\)}{Turbine diameter [$m$]}
\nomenclature{\(D_c\)}{Diameter of cylindrical deflector [$m$]}
\nomenclature{\(L_x\)}{Horizontal distance of deflector from turbine shaft [$m$]}
\nomenclature{\(L_y\)}{Vertical distance of deflector from turbine shaft [$m$]}
\nomenclature{\(k\)}{Turbulent kinetic energy [$m^2/s^2$]}
\nomenclature{\(y^+\)}{Dimensionless wall distance [-]}
\nomenclature{\(\omega\)}{Specific turbulent dissipation rate [$m^2/s^3$]}
\nomenclature{\(\Omega_z\)}{Z-Vorticity magnitude [1/s]}
\nomenclature{\(a\)}{Dilation parameter [-]}
\nomenclature{\(\xi_{\omega m}\)}{Shape Parameter [-]}
\nomenclature{\(g_1\)}{Upper limit of dilation parameter [-]}

\printnomenclature

\section{Introduction}
\label{}
The immense potential of clean energy sources such as wind, hydropower, geothermal, solar, and biomass energy has started to receive a lot of attention because of the growing energy needs of the world \cite{deshmukh2023renewable, HASSAN2024, wang2024remote}. The energy crisis has made it essential to find alternative energy sources and optimize the use of existing ones. The environmental effects and the availability problems associated with fossil fuels have only emphasized the urgent need to transition towards inexhaustible sources of energy. Harnessing these inexhaustible power sources can mitigate climate change, reduce air and water pollution, and ensure that future generations have access to clean energy. Wind energy is among the potential inexhaustible energy sources and is an increasingly developing alternative energy production method \cite{grujicic2010multidisciplinary}. The USA anticipates wind to provide 20\% of its total electricity production by 2030 \cite{thresher2008future}. This scenario can only be achieved by addressing the fundamental challenges identified in the context of harvesting wind energy. Approximately 10 million megawatts (MW) of wind energy is consistently available in the Earth's atmosphere, according to estimates \cite{mohamed2010optimization}. According to the Global Wind Energy Council (GWEC), countries investing in wind energy have collectively reached a record-high installed capacity of 906 GW in 2022. An additional 100 GW of new capacity is expected to have been added in 2023 \cite{gwec2023}. Despite the economic difficulties and resource limitations faced by wind turbine supply chains due to the pandemic, the GWEC reports an increment of 9\%  compared to the previous year in the wind power capacity in 2022. \par

Wind turbine design is an interdisciplinary research field that uses the fundamentals of fields such as turbomachinery, vibrations, aerodynamics, and material science to create efficient and sustainable energy solutions \cite{muscolo2012formulation}. By integrating multidisciplinary principles, researchers aim to optimize the performance of wind turbines, maximizing energy capture while minimizing structural fatigue and environmental impact. Large companies like Vestas and Siemens are actively trying to optimize wind turbine systems for efficient energy production. In academia, many researchers have proposed different optimization objectives, algorithms, constraints, and tools to improve wind turbine performance \cite{chehouri2015review}. 

Since the pioneering work of Betz in 1920 \cite{betz1920maximum}, substantial strides have been made in the field of wind turbine aerodynamics, especially their optimization. Betz \cite{betz1920maximum} found that the highest efficiency that a wind turbine may reach is 59.33\%. Wind turbines are classified into two main categories, namely vertical axis and horizontal axis, based on the orientation of their rotation axis \cite{schaffarczyk2020introduction}. Vertical axis wind turbines (VAWTs) have garnered significantly more attention than horizontal axis turbines due to their superior efficiency in low-wind conditions \cite{iliev2023investigation}. VAWTs further offer two main principles of operation: drag-based, and lift-based \cite{chen2015comprehensive}. For small-scale wind turbine projects, Manwell et al. \cite{manwell2010wind} pointed out that drag-based turbines offer lower construction costs than lift-based turbines. Dominy et al. \cite{dominy2007self} further suggested that drag-based turbines offer greater self-starting capabilities even in low wind speed conditions, unlike their lift-based counterparts. This makes drag-driven wind turbines aerodynamically more advantageous, especially on sites with variable wind speeds and directions. Even then, a major problem with the VAWTs is their low efficiency in converting wind energy into usable electricity \cite{talukdar2018parametric}. This is attributed to several factors, such as the geometry of the turbine, and the complex aerodynamic flow patterns they encounter. 

Currently, there is an increasing emphasis on enhancing the efficiency of Savonius turbines. Devised by Sigurd Johannes Savonius in the year 1929 \cite{johannes1929rotor}, the Savonius turbine features S-shaped blade design that rotates about a central axis. Recognizing the substantial potential of this turbine design, researchers are actively engaged in efforts to boost its efficiency. To achieve this goal, extensive research and development is going into modifying the blade design and profile to improve energy capture, as well as exploring wind deflectors as power augmentation devices to optimize the turbine's performance. Additionally, numerical techniques, including computational fluid dynamics (CFD), are being utilized to compute flow patterns and aerodynamic/hydrodynamic loads, thereby contributing to design improvements.

However, instead of relying solely on conventional numerical methods such as CFD, researchers are increasingly complementing them with novel techniques like surrogate modeling and meta-heuristic optimization algorithms. This reduces the number of simulations required and efficiently finds the optimal design parameters, thereby saving time and computational resources \cite{wu2024review, parekh2024surrogate, leng2024variable, lu2024surrogate}. Surrogate models are simplified mathematical representations derived from complex computer simulations or experiments that aid in approximating the behavior of the complex underlying system. Subsequently, meta-heuristic optimization algorithms are used to efficiently explore large solution spaces and obtain quasi-optimal solutions for optimization problems. These algorithms can be broadly categorized as evolutionary, nature-based, or swarm-intelligence algorithms and are often used when traditional optimization methods are impractical due to the non-linearity or complexity of the problem.

Various methods have been introduced in the literature for improving the efficiency of Savonius turbines. In their study, Kassem et al. \cite{kassem2018performance} altered the configurations of the end-plates for Savonius VAWTs and performed CFD studies with a 3 to 13 m/s wind speed range. Implementing both upper and lower end-plates, the study reported a 35\% increase in rotor power compared to using the turbine without end plates. Using Bezier curves and CFD simulations, Zemamou et al. \cite{zemamou2020novel} improved the Savonius turbine blade design, achieving a significant 29\% increase in power coefficient ($C_p$) over the standard VAWT. Xia et al. \cite{xia2022blade} utilized nature-based algorithms and surrogate models to optimize the blade shape of Savonius VAWTs, leading to a 7\% higher average torque coefficient than the traditional design at a TSR of 1. Furthermore, He et al. \cite{he2019performance} conducted CFD simulations and coupled the data with an Evolutionary Genetic Algorithm (GA) to improve the shape of the turbine blades and deflector placement for Savonius VAWTs, reporting a 34\% increase in the average $C_p$ with optimized blades compared to semi-circular blades. Additionally, the optimized deflector yielded a remarkable 95\% enhancement in the average $C_p$ compared to the baseline configuration of the turbine. In 2011, Golecha et al. \cite{golecha2011influence} undertook experiments to assess the influence of the placement of a rectangular deflector close to a Savonius hydro turbine. Optimizing the location of the deflector resulted in a 50\% augmentation in $C_p$, highlighting the pivotal role played by proper deflector positioning in boosting the turbine's performance.

The use of deflectors in Savonius VAWTs to improve power output has garnered significant attention among researchers \cite{he2019performance, singh2024hydrodynamic, kailash2012performance, tartuferi2015enhancement, al2019review, singh2023maximizing}. However, flow deflectors have a serious disadvantage in that they create strong turbulent wake regions downstream, which detrimentally effects the flow behavior, especially near the returning blade. This causes a negative torque output, resulting in a significant loss in the turbine's net power and torque output. The generation of strong vortices by flat plate deflectors induces flow instability, impacting the returning rotor and also limiting its application in a turbine array. Additionally, the interaction between the turbine rotors, and the vortices shed from the deflector, causes periodic loading on the turbine blades, which increases system fatigue \cite{fatahian2022innovative}. Researchers have shifted their focus to circular deflectors as a potential solution to address these challenges in Savonius VAWTs. The application of circular flow deflectors as power augmentation devices is intended to minimize wake regions, alleviate downstream vortices, and improve overall performance. Yuwono et al. \cite{yuwono2020improving} conducted experiments involving circular deflectors positioned ahead of the turbine's returning blade, resulting in a 12.2\% increase in the turbine's $C_p$ for a TSR of 0.65 compared to a conventional Savonius turbine. In 2018, Setiawan et al. \cite{setiawan2018numerical} introduced a circular cylinder next to the advancing blade of the Savonius turbine. The advancement of the blade and the circular cylinder's nozzle-like effect collectively facilitated blade acceleration, leading to an approximate 17.3\% enhancement in $C_p$, as per their numerical investigation. Furthermore, Singh et al. \cite{singh2023maximizing} utilized the Grey Wolf optimization algorithm and Kriging surrogate model to optimize Savonius turbine's performance with an upstream stationary cylindrical deflector and reported an enhancement of 34.2\% in the power output.

Even though a stationary circular deflector deflects the incoming flow towards the turbine's advancing blade, its stationary nature causes a significant flow separation, generating large and powerful vortices that destabilize the flow and adversely affect the returning blade. To counteract this, Fatahian et al. \cite{fatahian2022innovative} analyzed the impact of using upstream rotating cylindrical deflectors. In the study, the authors employed the commonly used heuristic method of finding the optimal parameters of the system and reported a 33.3\% peak increment in $C_p$ through the use of a deflector rotating at an angular velocity of 40 rad/s. Although their study is a good starting point for examining the influence of design parameters of the rotating deflector-Savonius turbine system, it does not provide an accurate representation of the impact of system parameters on performance as they have studied only the influence of one parameter by keeping the value of other parameters a constant and also have chosen a very small sample size of discreet parameter values.

To address the limitations of prior studies and introduce a robust and efficient framework for multi-parameter design optimization, the current work is structured around four main objectives:

\begin{enumerate}
    \item \textbf{Data Collection and Surrogate Model Training:} The first objective involves gathering data from CFD simulations and subsequently training a surrogate model.
    \item \textbf{Comparison of Different Surrogate Models:} The second objective involves comparing six different surrogate models and selecting the optimum one based on their accuracy at approximating the relation between $C_p$ and design parameters of the turbine-deflector system.
    \item \textbf{Implementation of Quantum-Based Salp Swarm Algorithm:} The third objective entails the utilization of the Quantum-based Salp Swarm Optimization (QSSO) algorithm to ascertain the optimal stationary cylindrical deflector-turbine configuration.
    \item \textbf{Comparison with Other Meta-heuristic Algorithms:} The fourth objective involves comparing the QSSO algorithm against nine other prominent meta-heuristic algorithms.
    \item \textbf{Determination of Optimal Rotational Velocity:} The final objective aims to determine the optimal rotational velocity of the cylindrical deflector, thereby achieving the maximum power output from the system.
\end{enumerate}

\section{Computational Model}
\label{}

\subsection{Parameterization}
\begin{figure} [h]
    \centering
    \includegraphics[width=1\textwidth]{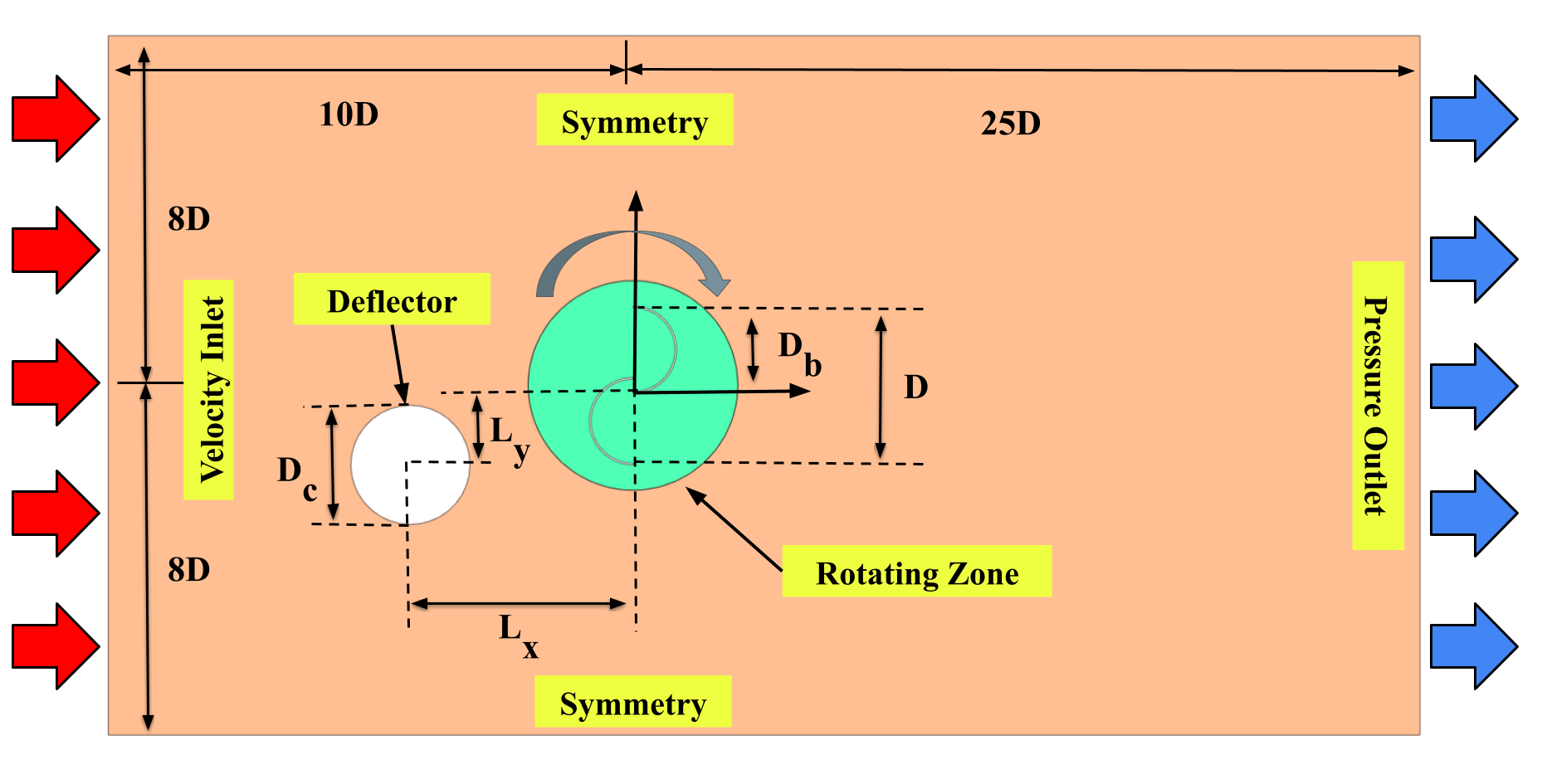}
    \caption{Flow domain schematic for the Savonius turbine-deflector system}
    \label{domain}
\end{figure}

In the current study, a drag-based Savonius wind turbine is utilized, incorporating a cylindrically shaped deflector situated upwind of the returning turbine blade. Figure \ref{domain} depicts a schematic representation of the turbine-deflector configuration, with the deflector placed just ahead of the returning blade. Dimensions of the system, viz. blade diameter ($D_b$) = $0.5$ $m$ and turbine diameter ($D$) = $0.909$ $m$ were sourced from the experiments of Yuwono et al. \cite{yuwono2020improving}, which noted that redirecting the air coming towards the front of turbine's returning blade can markedly enhance its performance. This phenomenon was seen to be contingent on four parameters: the diameter of the cylindrical deflector ($D_c$), the horizontal and vertical position of the deflector center from the center of the turbine shaft ($L_x$ and $L_y$ respectively), and the rotational velocity of the deflector ($\omega_d$). Previous investigations have also confirmed the sensitivity of these parameters to improve turbine performance \citep{yuwono2020improving, setiawan2018numerical}. Optimal system performance and efficiency necessitate limiting the parameter values within specific established limits derived from previous studies \cite{fatahian2022innovative, singh2023maximizing}. Therefore, for the current study, $L_x/D$ ranges from $1$ to $2$, the value of $L_y/D$ would be kept between $0.3$ and $1$, $D_c/D$ within $0.25$ and $1$, and $\omega_d$ would be maintained between $1$ and $50$ $rad/s$. While these constraints were primarily derived from the system's performance considerations, the design feasibility was also taken into account when determining these limiting values. \par

\subsection{Performance Parameters}
Wind turbines harness energy from the flowing wind. The theoretical energy present in the wind is formulated as: \begin{equation}
    P = \frac{1}{2} \rho  A V_\infty^3
\end{equation}
In the above equation, $\rho$ is the fluid density, $A$ is the swept area of the turbine and $V_\infty$ is the velocity of the incoming fluid. \par
However, wind turbines can extract only a fraction of this theoretically available power, the fraction being termed the power coefficient ($C_p$). Thus, the power extracted by the turbine is expressed as: \begin{equation}
    P_{extracted} = \frac{1}{2} \rho AV_\infty^3 \times C_p
\end{equation}

The power coefficient of any turbine cannot exceed the limit of 0.59, as specified by Albert Betz in 1920 \cite{betz1920maximum}. In other words, the maximum efficiency that can be attained by a wind turbine is 59.3\%.
A critical parameter, known as the tip speed ratio (TSR), characterizes the performance of a wind turbine and is the ratio of the turbine's tangential velocity to the velocity of the incoming fluid. This is given by:
\begin{equation}
    TSR (\lambda) = \frac{\omega_t R}{V_\infty}
\end{equation}

In the above equation, $\omega_t$ is the angular velocity of the turbine, $R$ is its diameter and $V$ is the velocity of incoming fluid.\par 
The TSR establishes a connection between the turbine's moment coefficient ($C_m$) and its power coefficient through the relationship:
\begin{equation}
    C_p = \lambda C_m
\end{equation}
The moment generated by the turbine blades can be computed using:
\begin{equation}
    M = \frac{1}{2} C_m\rho ARV_\infty^2
\end{equation}


\subsection{Computational Grid}
For the CFD simulations, the Savonius turbine was placed in a two-dimensional rectangular flow domain divided into two subdomains. To utilize the sliding mesh method, the inner domain was made to be the rotating zone, while the outer domain was made the stationary zone. The turbine blades were placed in the inner domain. To generate the grid for the simulations, the flow domain was partitioned into two zones: an inner section, comprising the deflector and the turbine blades, and the outer section. The inner section was meshed using unstructured tri-elements, whereas the outer section was meshed using structured quad-elements. \par 
This hybrid-meshing technique has two major benefits. Firstly, it prevents the formation of low-quality and high aspect ratio cells near the turbine and deflector walls. Secondly, it enables the use of a parametric framework that makes the simulation process more efficient. Numerous design points were computed in this investigation to attain the optimal solution. For each design point with new values of $D_c$, $L_x$, $L_y$, and $\omega_d$, a sequential procedure involving geometry update, remeshing, and solution solving is required. The hybrid-meshing approach facilitates the grid generator to concentrate primarily on modifying the grid only throughout the inner zone, while preserving the grid for the outer zone, resulting in a significant decrease in the overall computing time.
\par
\begin{figure} [h]
    \centering
    \includegraphics[width=1\textwidth]{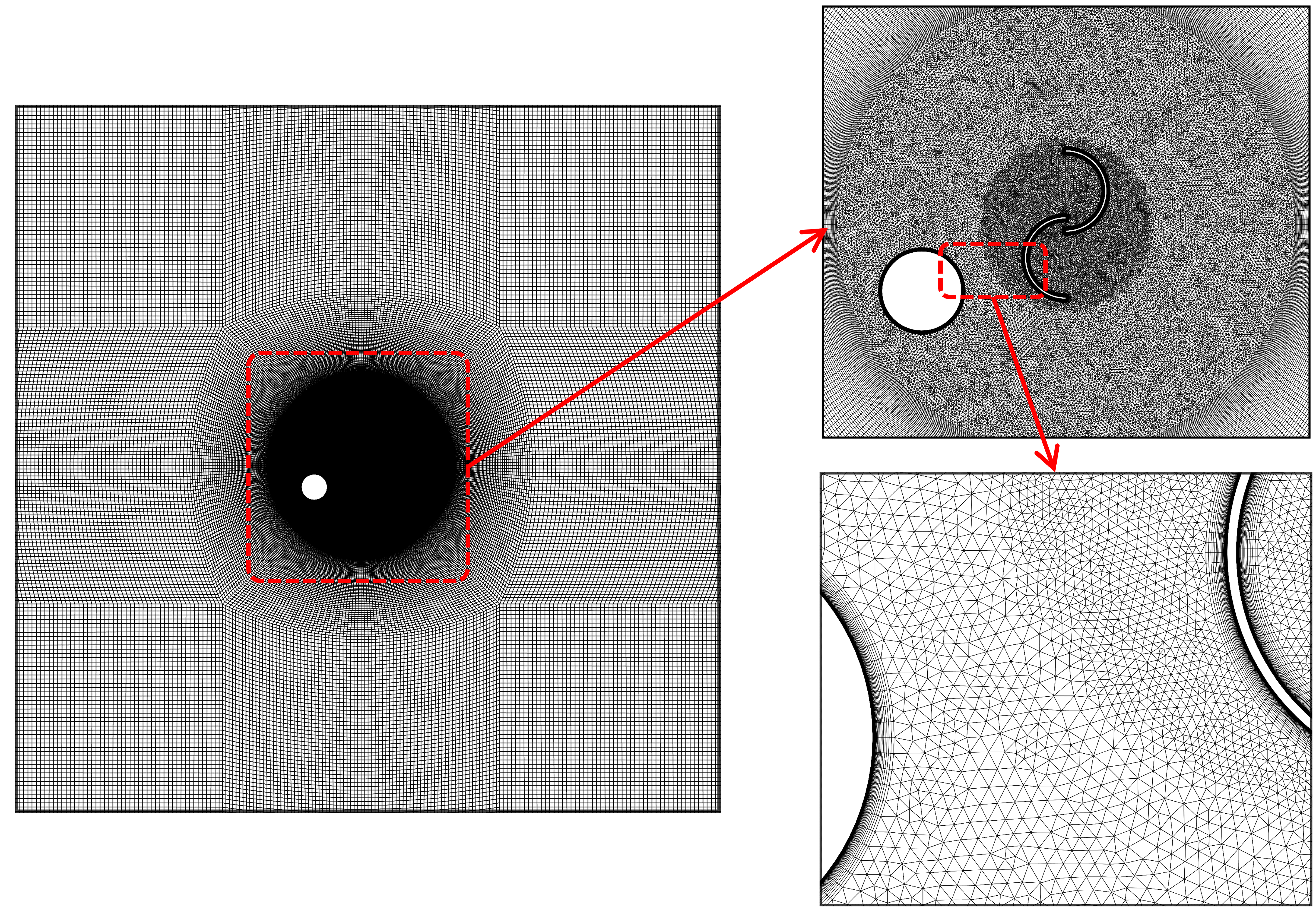}
    \caption{Computational Grid}
    \label{mesh}
\end{figure}

Figure \ref{mesh} depicts the grid used for the simulations. An extended domain is used in order to capture the vortices shed from the turbine-deflector system. To generate the hybrid mesh, the domain was split into 14 mesh blocks. A finer and more refined grid was implemented near the turbine and deflector walls, to accurately capture the complex physics of the turbulent viscous flow. While generating the near-wall cells, a $y^+$ value of less than 1 was maintained with 32 inflation layers wrapped around the deflector and turbine blades, having a first cell height of $5\times10^{-3} mm$, and a constant growth rate of 1.14. \newline

\subsection{Setup and Boundary Conditions}
Unsteady RANS simulations were conducted using ANSYS Fluent version 22.1 \cite{manual2009ansys} to generate training data for the surrogate model. A uniform freestream velocity of $7$ $m/s$ and constant pressure and density values were imposed at the inlet. The operating pressure was kept at $101325$ $Pa$ throughout the domain, while the density was kept at $1.225$ $kg/m^3$. Velocity boundary conditions at the turbine blades and the deflector walls were defined by imposing the no-slip condition. To simulate a fairly high turbulent inflow, the inlet turbulence intensity and the turbulent viscosity ratio were specified as $5\%$ and $10$, respectively. Dynamic viscosity of the fluid was taken as $1.7894 \times 10^{-5}$ $kg/m.s$. A pressure-outlet boundary condition was imposed at the domain outlet, which imposes a Dirichlet boundary condition for the static pressure (i.e., static gauge pressure = 0). To ensure proper convergence of the simulations, it was ensured that the residuals fall below $1 \times 10^{-6}$ for all cases.

\subsection{Numerical Model}
Ensuring the accuracy of the numerical model was crucial for training a reliable surrogate model based on the CFD simulation data. Since the Mach number (0.02) of the current study was much smaller than 0.3, a pressure-based solver was utilized and the flow was modeled as an incompressible fluid. The dimensional governing equations are given as follows: \par
  \begin{equation}
      \frac{\partial U_i}{\partial x_i} = 0
  \end{equation} 
  \begin{equation}
   \rho\frac{\partial U_i}{\partial t} + \rho\frac{\partial U_iU_j}{\partial x_j} = -\frac{\partial P}{\partial x_i} + \mu\frac{\partial^2 U_i}{\partial x_j^2} + \frac{\partial (-\rho\overline{U'_iU'_j})}{\partial x_j}
  \end{equation} 
  
where the term $-\rho\overline{U'_iU'_j}$ denotes the Reynolds stresses, P denotes the pressure, $\rho$ denotes the density of the fluid, and $U_i$ \& $U_j$ represent the velocity components. The turbulence closure model is an important aspect that affects the simulation accuracy by modeling the Reynolds stress term. Previous studies \citep{talukdar2018parametric, fatahian2022innovative, kassem2018performance, zemamou2020novel, he2019performance, golecha2011influence} show that the two-equation $k-\omega$ Shear-Stress Transport model given by Menter \cite{menter1994two} can represent the flow field around VAWTs accurately. Mathematically, this turbulence model can be written as follows:

\begin{equation}
    \frac{\partial (\rho k)}{\partial t} + \frac{\partial (\rho k U_i)}{\partial x_i} = \frac{\partial}{\partial x_j} \left[ \Gamma_k \frac{\partial k}{\partial x_j} \right] + G_k - Y_k
\end{equation}
\begin{equation}
    \frac{\partial (\rho \omega)}{\partial t} + \frac{\partial (\rho \omega U_i)}{\partial x_i} = \frac{\partial}{\partial x_j} \left[ \Gamma_\omega \frac{\partial \omega}{\partial x_j} \right] + G_\omega - Y_\omega + D_\omega
\end{equation}

In this context, $\Gamma_k$ and $\Gamma_\omega$ stand for the effective diffusivity of $k$ and $\omega$ respectively. The terms $G_k$ and $G_\omega$ represent the generation of $k$ and $\omega$ arising from velocity gradients, while $Y_k$ and $Y_\omega$ signify the dissipation of $k$ and $\omega$ respectively. The cross-diffusion term is denoted by $D_\omega$. \par
This model switches between the $k-\omega$ model of Wilcox \cite{wilcox2008formulation} near the wall, and the standard $k-\epsilon$ model \cite{spalding1974numerical} in far-field regions. This is done using a blending function and exploits the near-wall robustness of the Wilcox $k-\omega$ model while mitigating its freestream sensitivity by switching to the freestream-independent standard $k-\epsilon$ model. \par

Besides turbulence closure model, the simulation accuracy is also sensitive to the pressure-velocity coupling and discretization schemes. The COUPLED algorithm \cite{manual2009ansys} was utilized for pressure-velocity coupling in the simulations. Furthermore, a second-order linear upwind approach was used for the spatial discretization of turbulent kinetic energy, specific dissipation rate, pressure, and momentum. The gradient terms were discretized using the least squares cell-based scheme.


\subsection{Verification and Validation Studies}

For the verification study, first, a mesh independence test was conducted to ensure that the simulation results were insensitive to the grid resolution. A finer grid, although much more accurate, comes with a trade-off in computational efficiency. Hence, it was imperative to generate the ideal mesh, that had a balance between accuracy, and computational costs. Figure \ref{grid_independence} denotes the variation of $C_m$ acting on the turbine, as a function of the turbine's rotation angle for three levels of mesh resolution (70000, 130000, and 250000 elements). It is evident from the figure that negligible variation exists among the medium and fine grid resolutions. Further refinement would not affect the results, and hence, all the subsequent simulation results are reported using the medium mesh resolution.
\mycomment{\begin{figure} [h]
    \centering
    \includegraphics[width=1\textwidth]{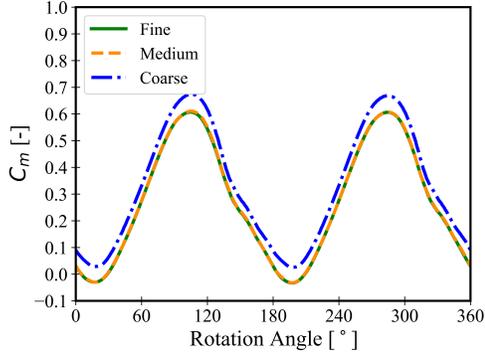}
    \caption{Grid Independence Test}
    \label{grid_independence}
\end{figure}
}

Further, a time-step dependence analysis was done to ensure that the results were independent of the temporal resolution. Just like mesh resolutions, a smaller time step tends to be more accurate but computationally expensive. It is important to employ a time step having a balance between an accurate solution and a low computational cost. Therefore, three time step sizes: $0.5^{\circ}$, $1^{\circ}$, and $2^{\circ}$ turbine rotation per time step were taken to investigate the effect of time step size on the solution accuracy. As seen from figure \ref{time_independence}, time step sizes smaller than $1^{\circ}$ would not affect the solution accuracy but the $2^{\circ}$ time step case leads to a phase difference in the results. Hence, all subsequent simulations are done using the time step corresponding to $1^{\circ}$ turbine rotation per time step.

\mycomment{\begin{figure} [H]
    \centering
    \includegraphics[width=1\textwidth]{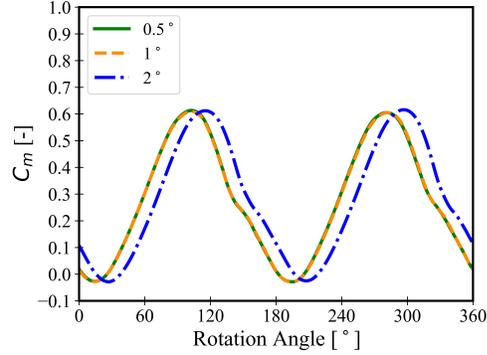}
    \caption{Time step Independence Test}
    \label{time_independence}
\end{figure}
}

\begin{figure}
\centering
\begin{minipage}{.5\textwidth}
  \centering
  \includegraphics[width=6.5cm]{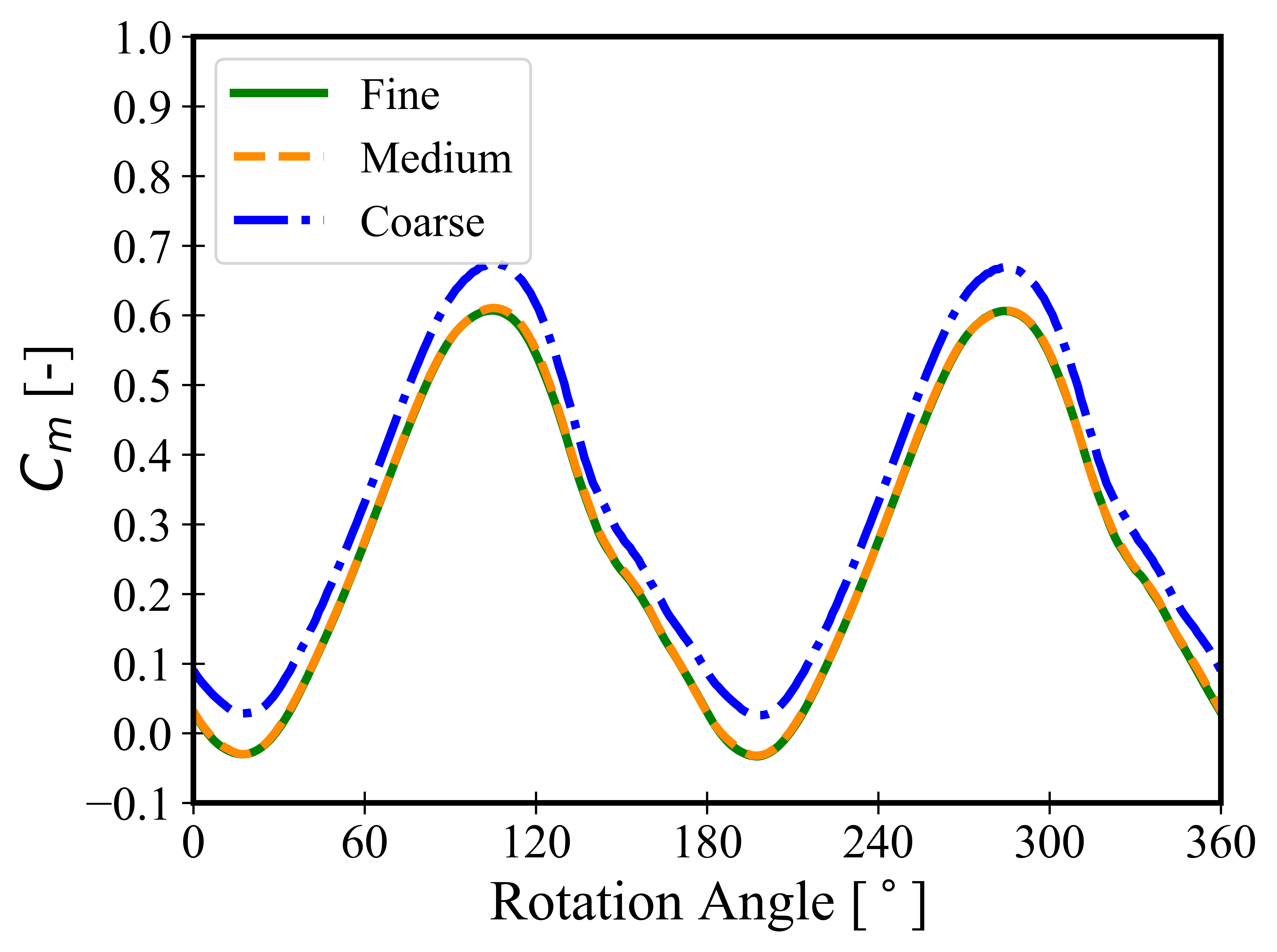}
  \caption{Grid Independence Test}
  \label{grid_independence}
\end{minipage}%
\begin{minipage}{.5\textwidth}
  \centering
  \includegraphics[width=6.5cm]{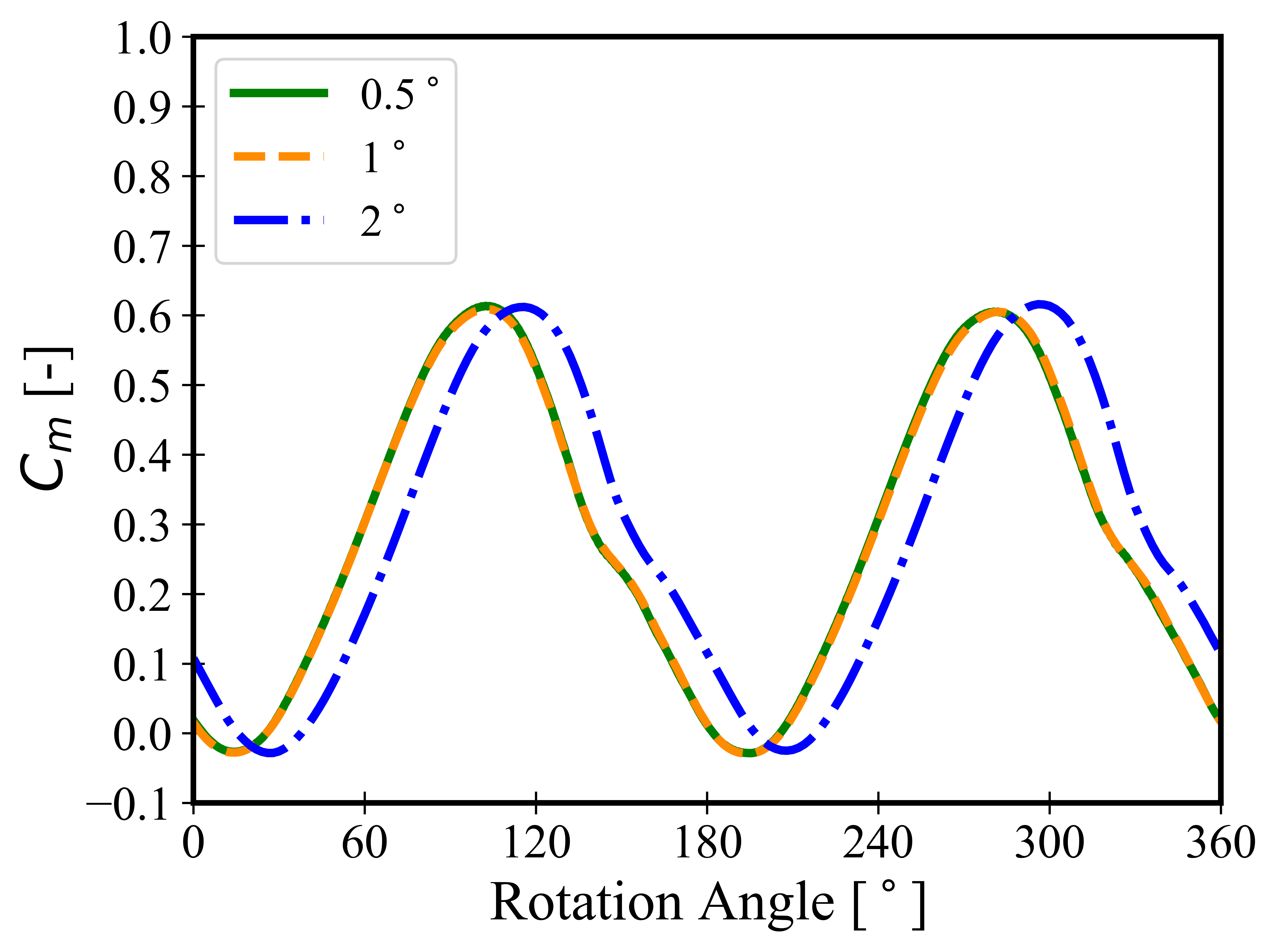}
  \caption{Timestep Independence Test}
  \label{time_independence}
\end{minipage}
\end{figure}

\begin{figure}
\centering
\begin{tabular}{cc}
  \includegraphics[width=65mm]{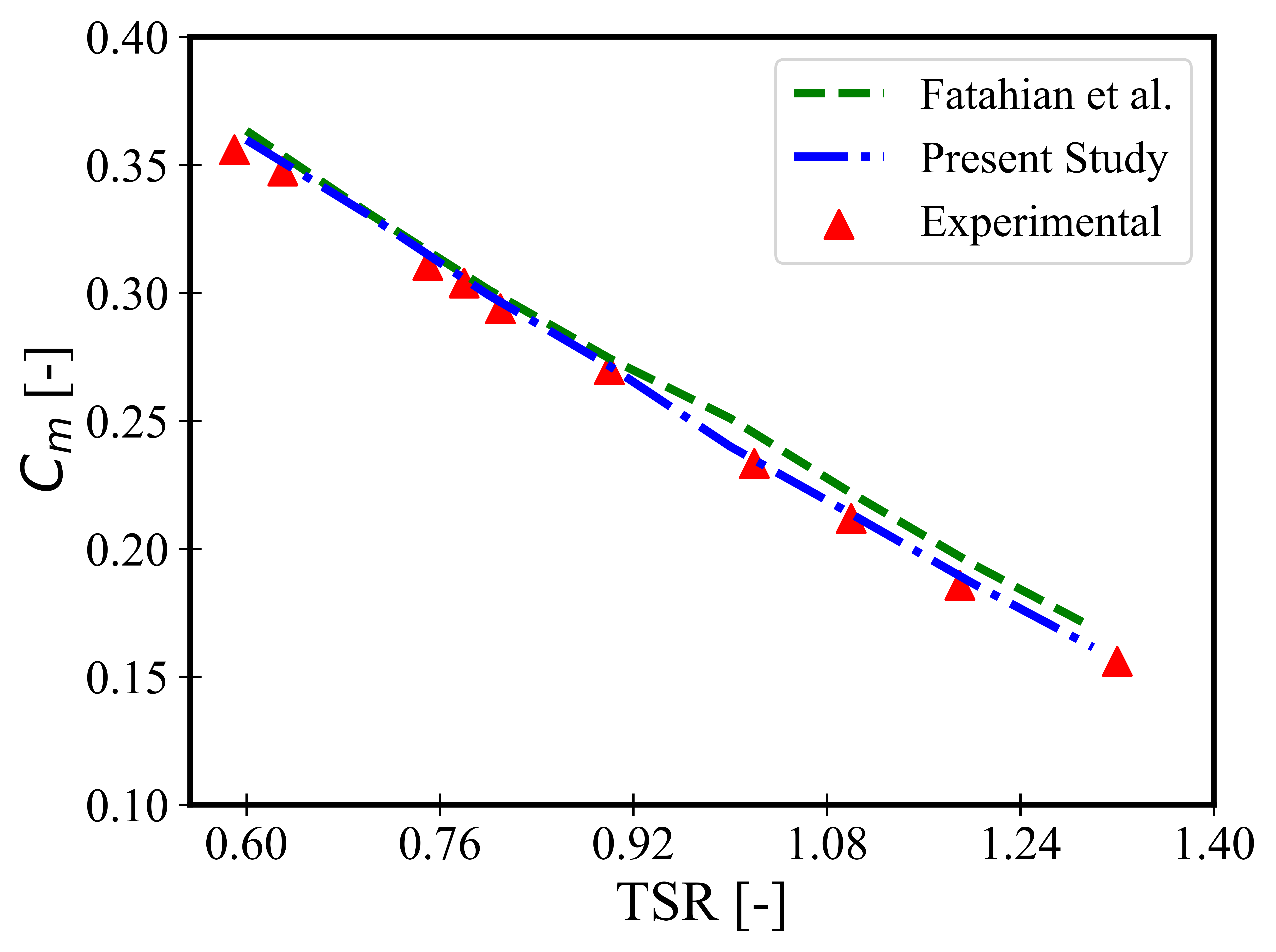} &   
  \includegraphics[width=65mm]{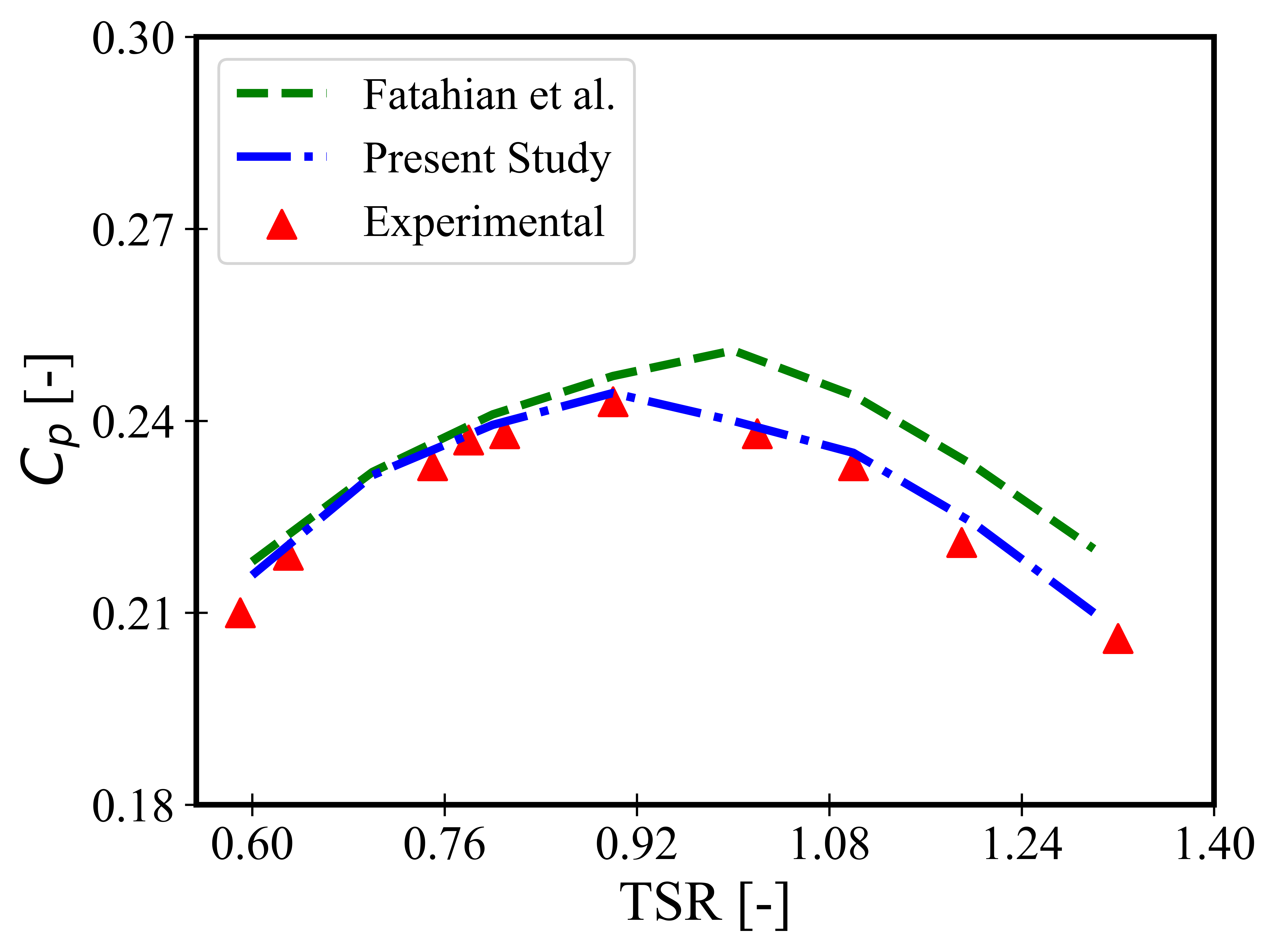} \\
(a) & (b) \\[6pt]
\end{tabular}
\caption{Comparison of the variation (a) $C_m$ and (b) $C_p$ with TSR generated using the current numerical model with the experimental results of Sheldahl et al. \cite{sheldahl1978wind} and numerical results of Fatahian et al. \cite{fatahian2022innovative}}
\label{Cp_Cm-validation}
\end{figure}

For the validation study, the results generated by the numerical model of the current study were compared with the experimental results of Sheldahl et al. \cite{sheldahl1978wind} and numerical results of Fatahian et al. \cite{fatahian2022innovative}. The results of the validation study are presented in Figure \ref{Cp_Cm-validation} in the form of plots representing the variation between $Cm$ and $C_p$ respectively with TSR for a conventional Savonius turbine. It is evident from Figure \ref{Cp_Cm-validation} that the results of the present study are in fair agreement with the experimental results and are also much closer to it in comparison to the numerical results of Fatahian et al. \cite{fatahian2022innovative}. \par

The results of these verification and validation tests confirm that the numerical model utilized in this study is in close agreement with experimental observations, and is independent of grid resolution and time step size. The validity of the present CFD model is hence substantiated.

\section{Optimization Framework}
\subsection{Sample Acquisitions}
The crucial role of sample acquisition in the optimization process cannot be overstated, serving as the fundamental basis for subsequent analyses. In the process of enhancing Savonius turbine performance through the strategic placement of a cylindrical deflector upstream of the blades, the primary goal is to collect pertinent data points that accurately capture the system's behavior. This involves conducting simulations across a range of operating conditions and design parameters. The application of Latin Hypercube Sampling (LHS) ensures a systematic distribution of data points throughout the design space, contributing to a comprehensive representation of the turbine's behavior.

A pivotal precursor to constructing a successful surrogate model is the generation of essential training data. The necessity for an equitable distribution of data points across the entire design space is paramount to guarantee the precision and dependability of the model. The adoption of Latin Hypercube Sampling (LHS), pioneered by McKay et al. \cite{mckay1992latin}, stands out as a widely embraced and effective technique for this purpose. This method meticulously selects data points, optimizing the coverage of the design space and establishing the foundation for developing a robust surrogate model geared towards streamlining the optimization of the system in question.

\subsection{Surrogate Modeling}
Design optimization problems often grapple with the high computational costs associated with a multitude of simulations \cite{wang2006review}. Surrogate models emerge as a cost-effective solution, constructing a response surface from a limited set of simulations to efficiently explore the design space and predict output variables. These models, trained on a subset of simulation data and validated using the rest, enable researchers to simulate complex systems more accurately with reduced computational resources. The versatility, simplicity, and efficiency of various surrogate models make them suitable for such investigations.

In the optimization context, the demand for a substantial number of computer simulations can be prohibitively expensive due to the need for significant computational resources \cite{xia2022blade}. Surrogate models play a pivotal role by acting as efficient approximations of the actual system behavior, streamlining the optimization process. These models use regression methods to infer response values for unknown design points based on known sample points, constructing an approximate representation with satisfactory prediction and fitting accuracy \cite{xia2022blade, tyagi2023novel, singh2023maximizing}. Unlike exclusive reliance on a single surrogate model, the present investigation adopts a comprehensive approach, utilizing various surrogate models to address the intricacies of significant nonlinear interactions between variables. This broader perspective enhances the adaptability and applicability of the optimization process.

\subsection{Quantum-Based Salp Swarm Optimization Algorithm}
To obtain the intended optimized design parameters, one must either maximize or minimize the objective function whose distribution across the design space is given by the most accurate surrogate model. To carry out this process, the present study employs meta-heuristic optimization algorithms. The inability of traditional algorithms to handle complicated problems and their slow convergence have given rise to quantum-inspired methods. A viable approach is the Quantum-Based Salp Swarm Optimization (QSSO) Algorithm, which takes inspiration from both the collective behavior of salp swarms and quantum mechanics concepts.

Salps, belonging to the Salpidae family, feature transparent barrel-shaped bodies akin to jellyfish, propelling themselves forward through water pumping \cite{madin1990aspects}. Despite their intriguing biological characteristics, research on salps is in its early stages, hindered by challenges in accessing their natural habitats and maintaining them in labs. Notably, salps exhibit a swarming behavior, forming chains in deep oceans. Although the purpose of this behavior is unclear, researchers speculate it enhances locomotion and foraging efficiency. Mirjalili et al. \cite{mirjalili2017salp} introduced the Salp Swarm Optimization (SSO) algorithm, drawing inspiration from the collective behavior of salps. Mimicking social interactions and movement dynamics observed in salp colonies, SSO is a metaheuristic optimization algorithm. Employing both exploration and exploitation strategies, SSO navigates solution spaces, refining promising regions. However, traditional SSO may face challenges in terms of convergence speed and exploration efficiency, motivating ongoing research to enhance its performance.

In addition to battling issues with time and space, the traditional SSO algorithm requires accuracy while locating solutions under nonlinear restrictions. Prioritizing efficiency improvement is crucial for addressing real-world problems. However, due to the restrictions inherent in biological algorithms and the lack of robust self-search conditions, numerous algorithms tend to encounter early convergence, which is also seen in the SSO Algorithm. Compared to other swarm intelligence algorithms, the SSO algorithm shows high competitiveness, a high value of accuracy, and practicability \cite{abbassi2019efficient}. In order to prevent the algorithm from falling into the local optimal solution Faris et al. \cite{faris2018efficient} proposed the binary salp swarm algorithm to enhance the exploration, and utilize the transfer functions and crossover operator to tackle Feature selection problems. In their published research, Sahu et al. \cite{sahu2018improved} concentrated on introducing a type-II fuzzy PID controller aimed at maintaining both frequency and tie-line power at their nominal values amidst various uncertainties. To determine optimal gain values, they applied a meta-heuristic improved Salp Swarm Optimization (I-SSO) algorithm. The study demonstrated that the I-SSO algorithm outperformed other meta-heuristic algorithms, as evidenced by superior dynamic response results.

The present work uses the Quantum-based Salp Swarm Optimization (QSSO) algorithm to determine the best location for a cylindrical deflector near the Savonius turbine. Chen et al. \cite{chen2019qssa} have developed a new technique that smoothly incorporates key notions from quantum mechanics, namely Elite Opposition-based and Wavelet Mutation, into the conventional SSO algorithm. By using quantum-inspired techniques, the QSSO algorithm seeks to improve optimization by facilitating effective exploration of the solution space and convergence towards an ideal solution. The QSSO method has the potential to improve exploration and tackle the problems of premature convergence and solutions locked in local optimum by utilizing the combined abilities of quantum mechanics and swarm intelligence.

\begin{algorithm}[h]
\SetAlgoLined
Set the population size $N$ and max iterations $T$ \par
Randomly initialize the positions of salps $\mathbf{x}_i$ $ (i=1,2,.., N)$ according to $ub_j$ and $lb_j$ \par
\textbf{$t$ $=$ $0$} \par
Calculate the fitness of each salp and set F as the best search agent \par
Update the positions of the leader and follower salps by Eq.\ref{conventional} \par
\textbf{$t$ $=$ $t$ $+$ $1$} \par
\While{$t$ $<$ $T$}{
    \If{$i$ $<$ $N/2$}{
        Update the position of leading salp using Eq.\ref{Q-update}
    }\par
    \Else{
        Calculate the position of follower salps using reverse elite Eq.\ref{elite} \par
        Compare the reverse elite solution and normal solution, update the Follower salp position with best fitness 
    }\par
    \textbf{end if} \par
    Random individual is selected and it's position is updated using wavelet mutation Eq.\ref{mutation} \par
    Correct the each salp that go beyond the upper and lower bounds \par
    Recalculate the fitness of each salp \par
    Update the best position of the food source (F) \par
}
\textbf{end while} \par
Next generation until stopping \par
\Return F
\caption{Quantum Salp Swarm Algorithm}
\label{algo}
\end{algorithm}

\textit{Working of QSSO algorithm}: Let the position vector of each salp be an $N$ $\times$ $D$ dimension array, where $N$ represents the number of salp search agents and $D$ is the dimension of continuous solution space. This vector can be represented in the form of a matrix, as shown in Eq.\ref{Matrix} :

\begin{equation}
    X_z^y = \left[
            \begin{array}{cccc}
            X_1^1 & X_1^2 & \cdots &X_1^D\\
            X_2^1 & X_2^2 & \cdots &X_2^D\\
            \vdots & \vdots & \cdots &\vdots\\
            X_N^1 & X_N^2 & \cdots &X_N^D
            \end{array}
            \right]
    \label{Matrix}
\end{equation}

The following equation is used by the conventional SSO algorithm to update the leader salp's position :

\begin{equation}
    X_z^1 =\begin{cases}
            F_z + c_1((ub_z - lb_z)c_2 + lb_z) & c_3 \ge 0.5 \\
            F_z - c_1((ub_z - lb_z)c_2 + lb_z) & c_3 < 0.5
            \end{cases}
    \label{conventional}
\end{equation}

with 

\begin{equation}
    c_1 = 2e^{-(l/L)^2}
    \label{c1}
\end{equation}

Where the position of the first generation leader salp of the population in the $z_{th}$ dimension is given by $X_z^1$, $l$ represents the current iteration of the algorithm and the maximum number of iterations is represented by $L$. In contrast, in the QSSO algorithm, the leading salp position update strategy takes inspiration from the Monte Carlo method which is used in quantum mechanics to convert the state of an individual from a quantum state to a classical state. Therefore, the strategy is described by the following Equation:  

\begin{equation}
    X_{z,g+1}^1 =\begin{cases}
            b_{z,g}^1 + c_1(M - X_{z,g}^1).\ln(1/c_2) & c_3 \ge 0.5 \\
            b_{z,g}^1 - c_1(M - X_{z,g}^1).\ln(1/c_2) & c_3 < 0.5
            \end{cases}
    \label{Q-update}
\end{equation}

where $b_{z,g}^1$ is the global optimum value of the $g_{th}$ generation, $M$ represents the average of local optimum values. The introduction of quantum mechanics effectively improves the mobility of salps. Different algorithms use different approaches to effectively search the solution space for the global optimum. Here, QSSO uses wavelet mutation and reverse elite learning to efficiently reach the global optimum. Reverse elite solution is used to calculate the position of follower salps and can be described by the following equation:

\begin{equation}
    X_z^y(elite) = c(ub_z + lb_z) - X_z^y(normal)
    \label{elite}
\end{equation}

where $c$ is a random number between $0$ and $1$, $X_z^y(normal)$ and $X_z^y(elite)$ are the positions of follower salp and their reverse elite solutions, respectively. Fitness values of both are calculated and the best one is chosen for the next generation while preserving the reverse elitism to some predefined criteria. In this way reverse elite strategy provides a more appropriate and robust candidate solution. Subsequently, Wavelet mutation is introduced to maintain population diversity and it works by giving each follower salp of the population a chance to mutate which is controlled by the parameter known as the probability of mutation where a random number is generated between 0 and 1 if it is less or equal to some predefined number and then mutation will take place. If $\overline{X}_z^y(t)$ is the selected salp, it will mutate according to the equation:

\begin{equation}
    \overline{X}_z^y(t) =\begin{cases}
            X_z^y(t) + \sigma \times (ub_z - X_z^y(t)) & \sigma > 0. \\
            X_z^y(t) - \sigma \times (X_z^y(t) - lb_z) & \sigma \le 0
            \end{cases}
    \label{mutation}
\end{equation}

with 

\begin{equation}
    \sigma = \frac{1}{\sqrt{a}}e^{\frac{-(\frac{x}{a})^2}{2}}\cos({5(\frac{x}{a})})
    \label{sigma}
\end{equation}

where $a$ is a dilation parameter that changes in each iteration to achieve fine-tuning. This dilation parameter is defined as:

\begin{equation}
    a = e^{-\ln{g_1}\times(1-\frac{l}{L})^{\xi_{\omega m}}+\ln{g_1}} 
    \label{dialtion}
\end{equation}

\begin{figure}
  \centering
      \includegraphics[width=14cm]{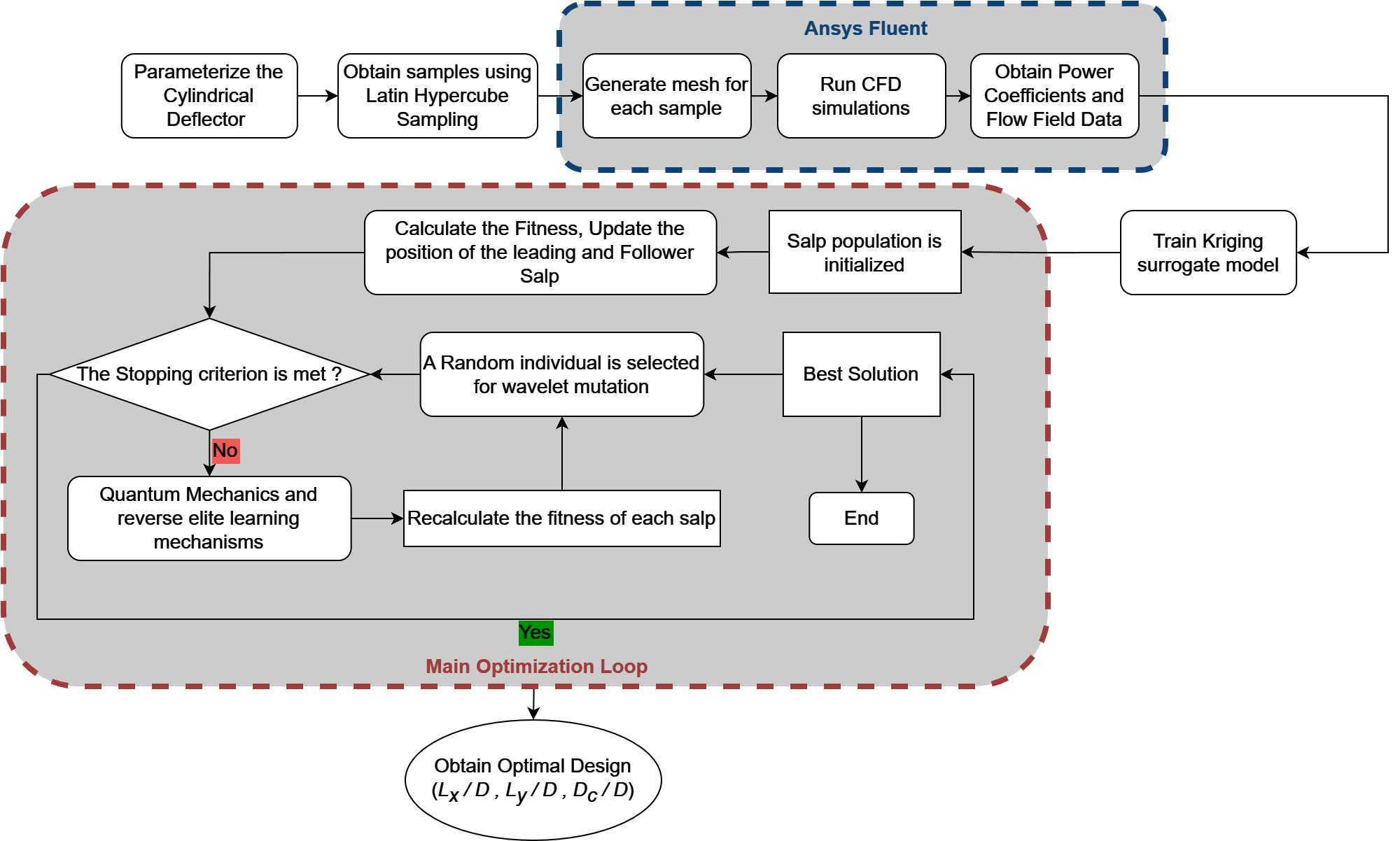}
      \caption{Flowchart of Optimization Framework}
      \label{QSSA}
\end{figure}

Here $\xi_{\omega m}$ is the shape parameter of this function and $g_1$ is the upper limit of dilation parameter $a$. By fine-tuning these two parameters one can get the peak performance out of QSSO. As done in reverse elite strategy, among the fitness of the mutated individual and non-mutated individual, the best one is selected for the next generation. The pseudo-code of the QSSO algorithm is provided in Algorithm 1 and the summary of the proposed optimization framework is given in Figure \ref{QSSA}.

\section{Results and Discussion}
\label{}

\subsection{Kriging Surrogate Model}
\begin{figure}[h]
\begin{tabular}{cc}
  \includegraphics[width=75mm]{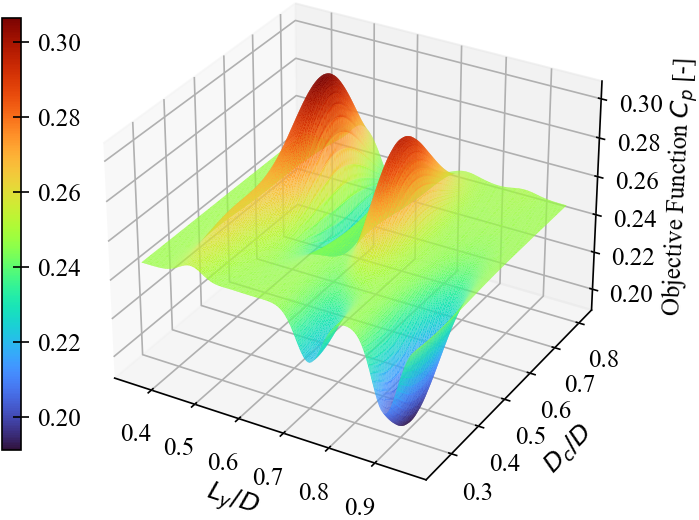} &   \includegraphics[width=75mm]{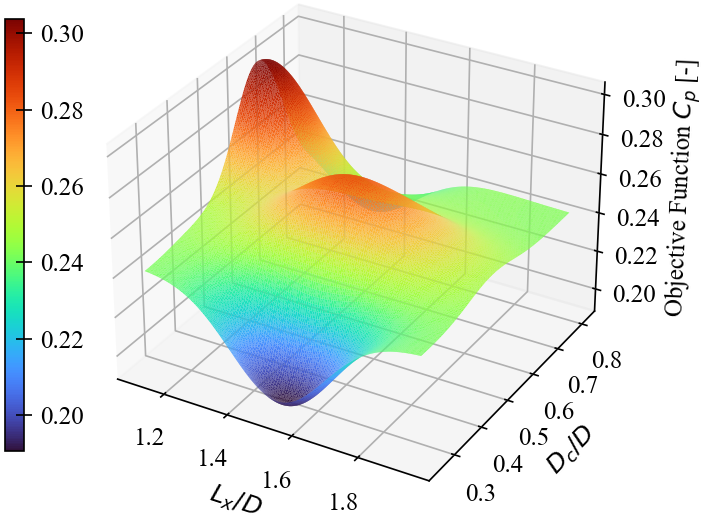} \\
(a) & (b) \\[6pt]
\multicolumn{2}{c}{\includegraphics[width=75mm]{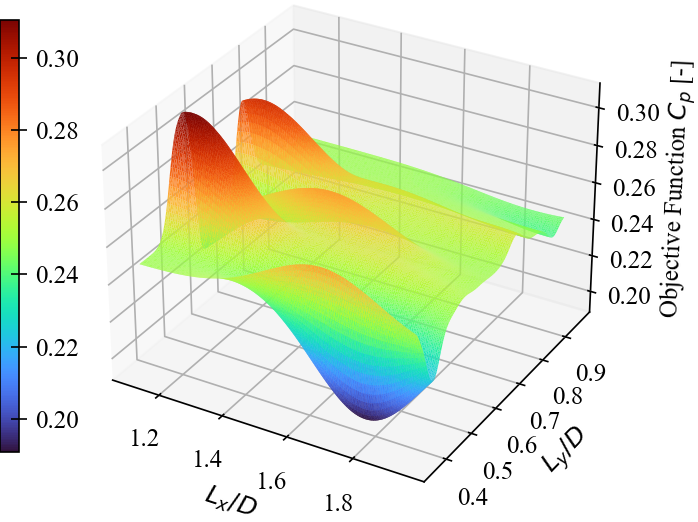} }\\
\multicolumn{2}{c}{(c)}
\end{tabular}
\caption{Response surfaces generated using Kriging model}
\label{Responses}
\end{figure}

A total of sixty sample points were generated for training the surrogate model, and CFD simulations were performed at a TSR of 0.9. The decision was made in light of the results of a previous study by Fatahian et al. \cite{fatahian2022innovative}, which showed peak values of $C_p$ at a TSR of $0.9$. Out of the $60$ sample points, $45$ were selected for the training set, while the others were used for validating the surrogate model. Six different surrogate models were tested on the numerically generated dataset, with the Kriging model displaying superior performance, resulting in an $R^2$ score of $0.943$. Three different cases were considered to find how the design parameters affected the objective function. One of the three design parameters ($L_x/D$, $L_y/D$, and $D_c/D$) was held constant in each case. Figure \ref{Responses} shows the response surfaces produced by the surrogate model for each instance. A strong non-linear association between the objective function and the design parameters $L_x/D$, $L_y/D$ and $D_c/D$ can be seen in Figure \ref{Responses}. According to the response surface analysis, the ideal solutions can be found for $L_x/D$ values of $1.0$ to $1.1$, $L_y/D$ values of $0.5$ to $0.6$, and $D_c/D$ values of $0.5$ to $0.7$. These results offer useful information for maximizing the circular cylinder design parameters by identifying the precise $L_x/D$, $L_y/D$, and $D_c/D$ ranges that greatly increase the Savonius turbine's power coefficient $(C_p)$ and provide clarity over the dependence of $C_p$ on the value of the design parameters.

\subsection{Performance Against Other Surrogate Models}
The performance of the Kriging surrogate model is quantitatively evaluated by comparing its $R^2$ score with that of other models. The following models were employed for the comparative analysis:
\begin{itemize}
    \item 
    Kriging: Kriging is an interpolating model that involves linearly adding a known function to a stochastic process realization, named after its pioneer Danie G. Krige.  
    \item
    KPLS: KPLS is a fast, accurate variant of the Kriging model that minimizes hyperparameter estimation through the use of the effective PLS (Partial Least Squares) method. It builds a PLS-based kernel that is well-known for maximizing the variation between input and output variables, making it perfect for high-dimensional issues \cite{bouhlel2016improved}.
    \item 
    KPLSK: Based on KPLS, the KPLSK model is built in two primary stages. First, the hyperparameters are estimated using KPLS. The probability function of a typical Kriging is locally optimized using these estimated hyperparameters as a starting point in the second step, which involves transforming them back into the original space. With KPLS, an educated initial guess is made for the hyperparameters. Gradient-based optimization is then applied using traditional Kriging kernels.
    \item 
    Inverse-Distance Weighting (IDW): A weighted average of the sampling points is used to compute the unknown points in the inverse distance weighting (IDW) model, which is an interpolation technique.
    \item 
    Radial Basis Function: RBF approximates complex functions with a combination of simpler functions. Previous research has examined how well the Radial Basis Function (RBF) has defined significant nonlinear interactions between variables \cite{xia2022blade, hou2014optimization, tyagi2023novel}. Its effectiveness, versatility, ease of use, and simplicity have made it a suitable stand-in model for a variety of research questions.
    \item 
    Support Vector Machine (SVM): Support Vector Machine (SVM) is typically used for classification tasks. They work very well for high-dimensional data and permit the classification of data that does not have a linear correspondence.
\end{itemize}

\begin{figure}[H]
\centering
\begin{tabular}{cc}
  \includegraphics[width=0.45\linewidth]{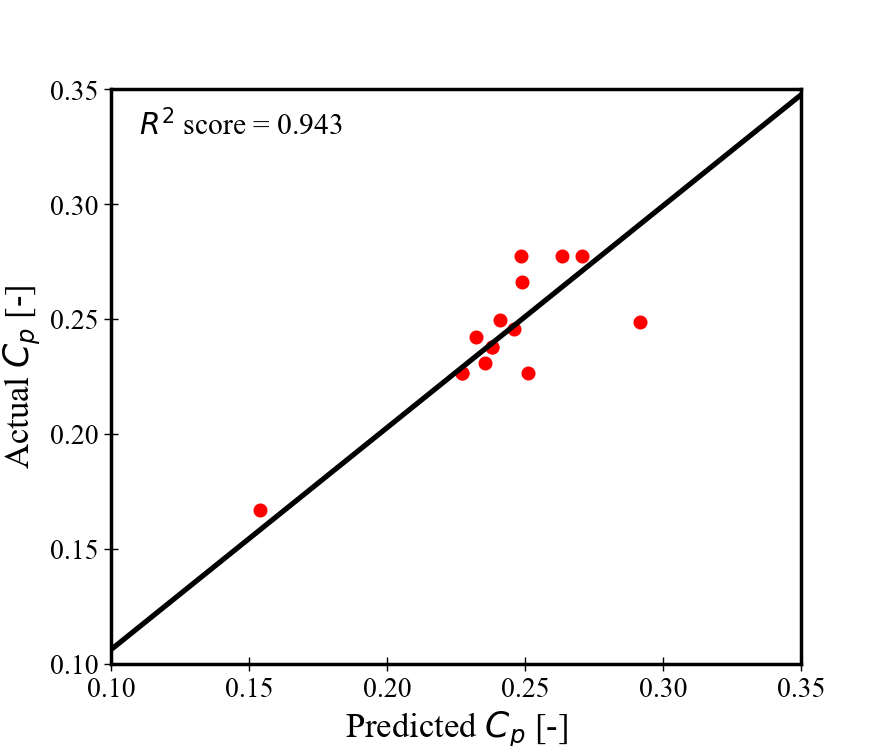} &   
  \includegraphics[width=0.45\linewidth]{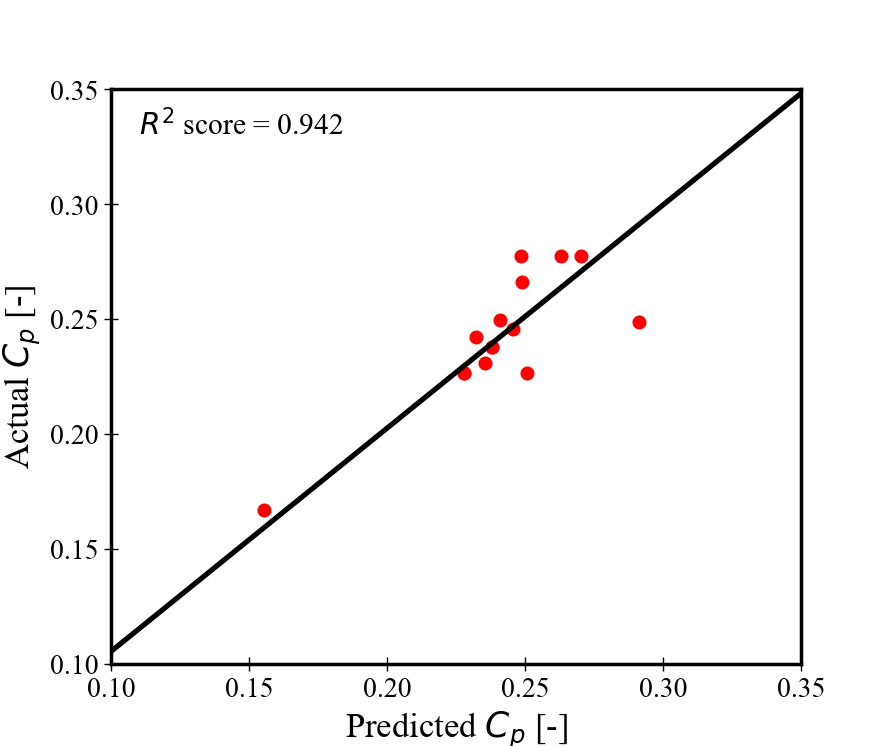} \\
  (a) & (b) \\[6pt]
  \includegraphics[width=0.45\linewidth]{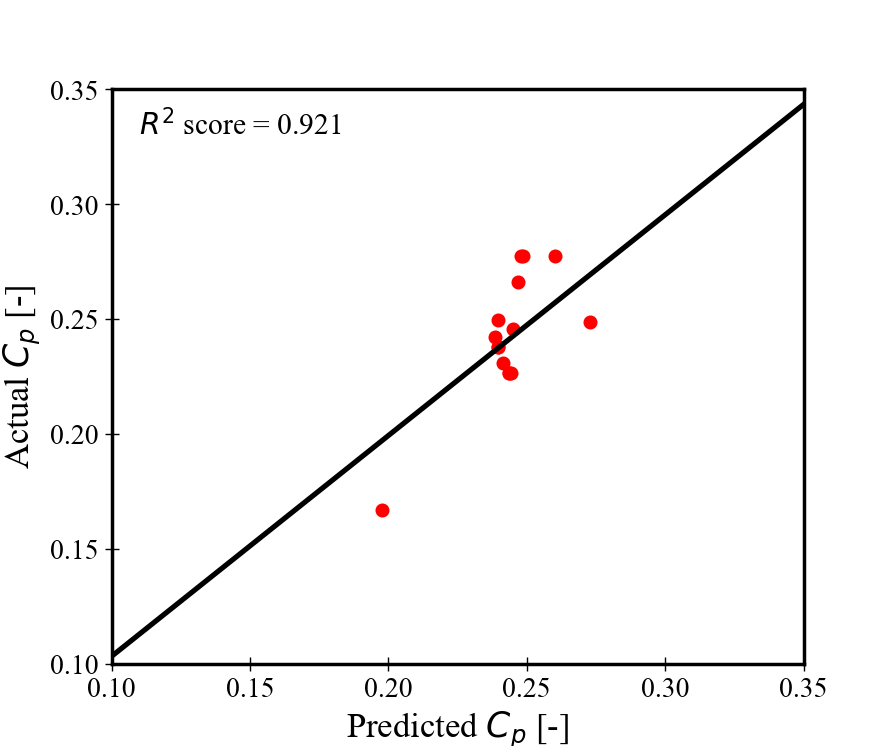} &
  \includegraphics[width=0.45\linewidth]{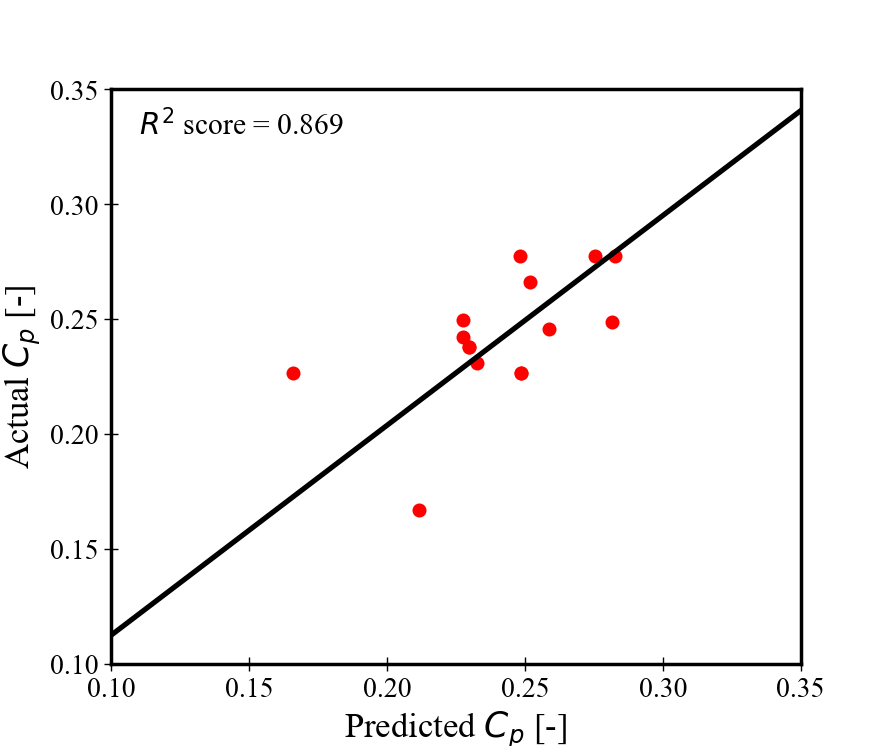}\\
  (c) & (d) \\[6pt]
  \includegraphics[width=0.45\linewidth]{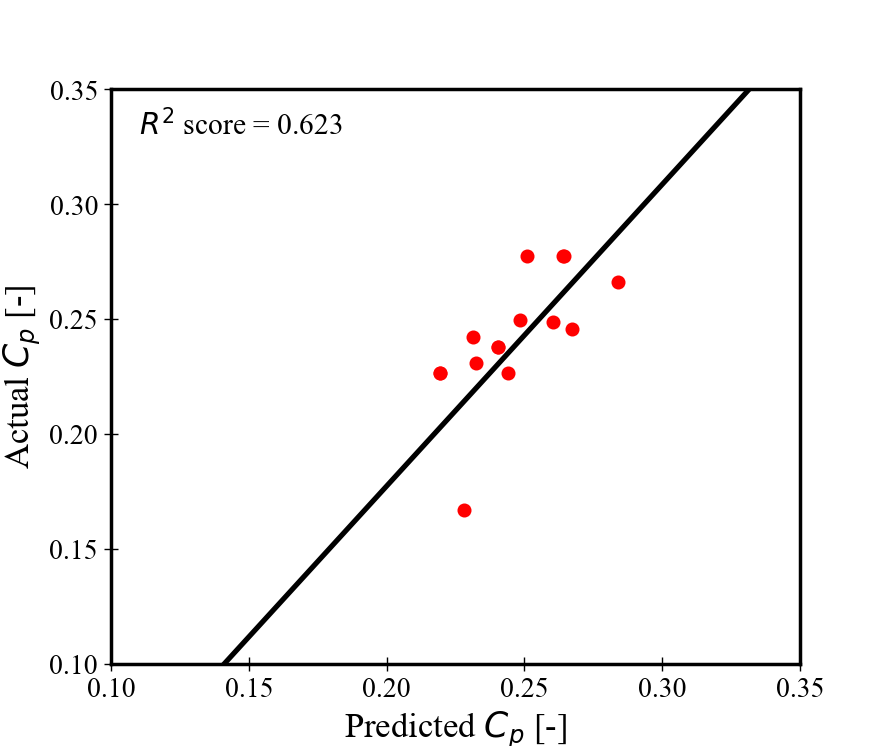} &
  \includegraphics[width=0.45\linewidth]{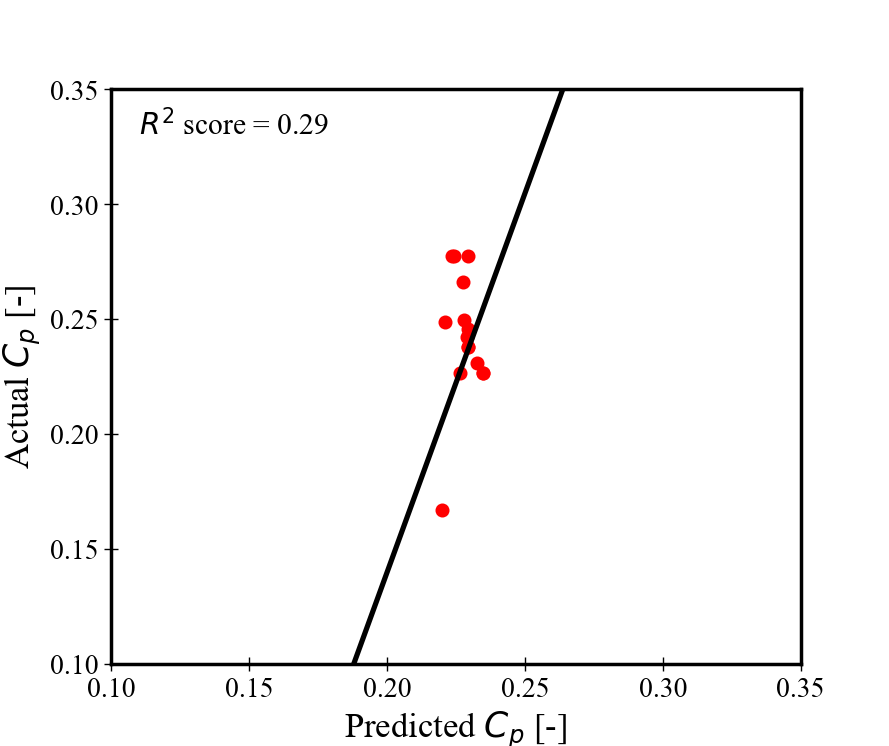}\\
  (e) & (f) \\[6pt]
\end{tabular}
\caption{Best Fit line using (a) Kriging, (b) KPLSK, (c) KPLS, (d) IDW, (e) RBF, and (f) SVM surrogate model}
\label{convergence}
\end{figure}

Figure \ref{convergence} shows the comparison of the different surrogate models at the 15 validation sample points. It can be observed that the KPLSK model trails closely behind the Kriging model, which has the greatest $R^2$ score of $0.943$ and the optimum best-fit line. Furthermore, as the scatter plot illustrates, the KPLS and IDW models do not correlate as well as the Kriging and KPLSK models, and the SVM and RBF models have a large deviation from the actual value in their results. Therefore, the Kriging model is chosen for the optimization of the design parameters. The above models are implemented using the open-source Surrogate Modeling Toolbox Python package \cite{saves2024smt}.

\subsection{QSSO Algorithm Hyper-parameters}
\begin{figure}[h]
\begin{tabular}{cc}
  \includegraphics[width=70mm]{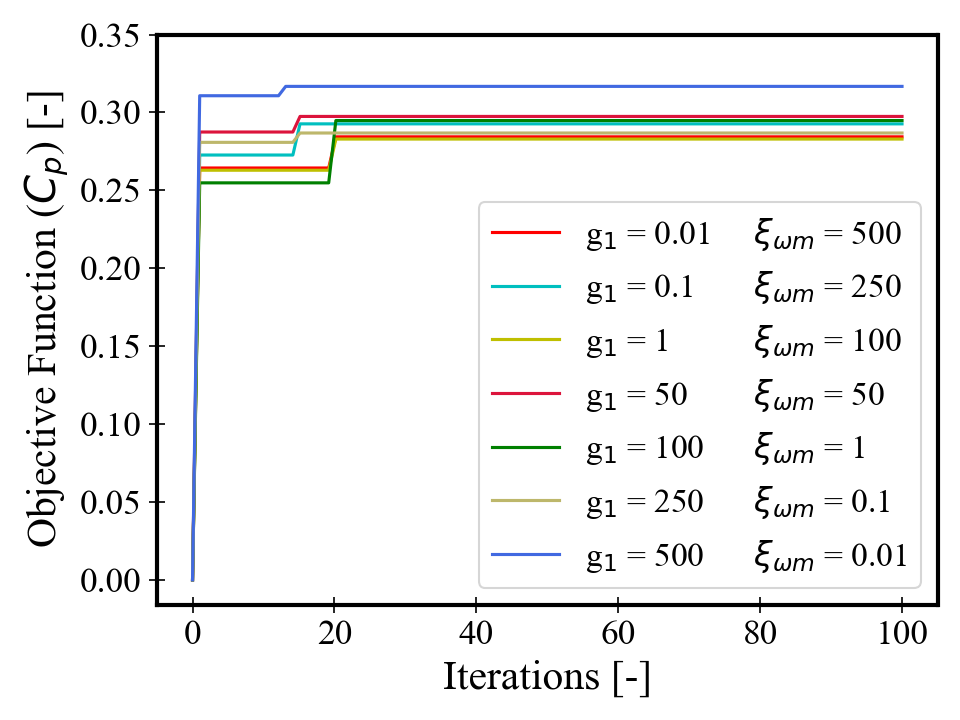} &   
  \includegraphics[width=70mm]{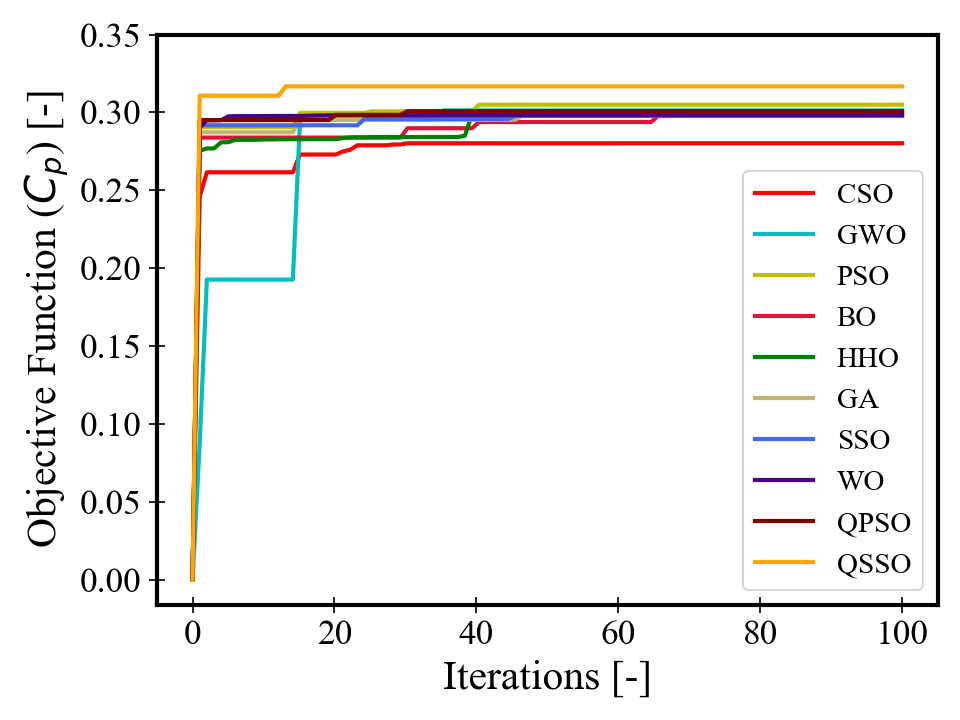} \\
(a) & (b) \\[6pt]
\end{tabular}
\caption{(a) Performance of QSSO algorithm at different values of $g_1$ and $\xi_{\omega m}$ (b) Performance of other metaheuristic algorithms compared to QSSO algorithm}
\label{Hyperparameter_Algo_Comparison}
\end{figure}

An in-house code for the QSSO algorithm was employed to maximize the objective function, i.e. $C_p$. The influence of the dilation parameter used in the wavelet mutation of the QSSO algorithm is presented in Figure \ref{Hyperparameter_Algo_Comparison}(a). This parameter, as the equation shows, is made up of two hyper-parameters: $g_1$, which is the dilation parameter's maximum value, and $\xi_{\omega m}$, the shape parameter. The convergence of the optimization algorithm was assessed for several values of (g1, $\xi_{\omega m}$), namely $(0.01, 500)$, $(0.1, 250)$, $(1, 100)$, $(50, 50)$, $(100, 1)$, $(250, 0.1)$ and $(500, 0.01)$. It is evident from the findings shown in Figure \ref{Hyperparameter_Algo_Comparison}(a) that QSSO performs best when $g_1$ and $\xi_{\omega m}$ are set to $500$ and $0.01$ respectively. A comparison of this setup to other combinations of $g_1$ and $\xi_{\omega m}$ values reflects that the algorithm converges faster and yields the best possible solution.

\subsection{Performance Against Other Algorithms}
\label{Sec-algo-comp}
The results obtained by the QSSO algorithm are quantitatively compared against other optimization algorithms to objectively assess its performance. For optimal computational efficiency, the population size or in other terms the number of search agents was set to 50. The following meta-heuristic algorithms were used for comparison: 
\begin{itemize}
    \item 
    Particle Swarm Optimisation (PSO): PSO is a method that utilizes the swarm intelligence of bird flocks to optimize systems. It is inspired by the social behavior of fish and birds. Its effectiveness and simplicity have made it a useful tool for resolving optimization issues \cite{kennedy1995particle}.
    \item 
    Quantum-Based Particle Swarm Optimization (QPSO): It is a variation of PSO that combines trajectory analysis from the original PSO with probabilistic components influenced by quantum mechanics \cite{heidari2019harris}.
    \item 
    Harris’ Hawk Optimizations (HHO): HHO is a novel population-based, nature-inspired optimization algorithm. Its basis is found in the way Harris's hawks hunt together, with multiple hawks working together to approach prey from various angles \cite{mirjalili2016whale}.
    \item 
    Whale Optimization (WO): Similar to HHO, WO is a novel swarm-based optimization technique that addresses complex optimization problems by simulating coordinated efforts and natural instincts inspired by humpback whale hunting techniques \cite{mirjalili2016whale}. 
    \item 
    Bat Optimization (BO): It is a novel optimization technique derived from nature that mimics the echolocation behavior of bats to effectively explore and identify optimal solutions in search spaces \cite{yang2010new}. 
    \item 
    Grey Wolf Optimization (GWO): GWO is an optimization algorithm influenced by nature. It draws influence from the hunting habits and social hierarchy of grey wolves \cite{mirjalili2014grey}.
    \item 
    Cuckoo Search Optimization (CSO): CSO is a nature-based algorithm that is inspired by cuckoo bird breeding behavior. It uses a population-based method to optimize problems, focusing on exploration-exploitation trade-offs \cite{tyagi2023novel}.
    \item 
    Salp Swarm Optimization (SSO): SSO algorithm takes its cues from the salp's collective behavior. Through both exploration and exploitation tactics, SSO navigates solution spaces, honing down on potential regions while mimicking the social interactions and movement patterns seen in salp colonies \cite{mirjalili2017salp}. 
    \item 
    Genetic Algorithm (GA): GA is an evolutionary-based metaheuristic algorithm, and it derives inspiration from the evolution of species. With the use of mutation and crossover techniques, it achieves the optimal solution.
\end{itemize}

As evident in Figure \ref{Hyperparameter_Algo_Comparison}(b), the QSSO algorithm demonstrates the highest percentage of improvement in the objective function, closely followed by SSO. The exploration capabilities of the QSSO, which allow it to produce new solutions and successfully escape local optima, are responsible for its superior performance. The exploration capacity of QSSO is further boosted by the use of reverse elite learning and wavelet mutations. On the other hand, even with high rates of convergence, QPSO and GWO fall short of SSO and QSSO in terms of reaching the best results; this is also the case for GA. Depending on the state-of-the-art solutions at the moment, HHO and PSO are vulnerable to trapping in local optima. Because they rely on adaptive mobility and random search, respectively, BO and WO do not have as great exploratory skills in comparison to the other investigated algorithms. CSO struggles to get strong results compared to QSSO, possibly because it relies on cuckoo search processes that can be problematic in some optimization settings, restricting its ability to escape local optima.

\subsection{Torque and Power Analysis}
Through an extensive optimization study, utilizing several state-of-the-art algorithms, the optimum geometric parameters of the turbine-deflector system were calculated. As presented in subsection \ref{Sec-algo-comp}, the system parameters provided by the QSSO algorithm gave the maximum improvement in the value of $C_m$ and $C_p$ in comparison to the baseline turbine without an upstream deflector. For the optimum configuration, the system parameters are $L_x/D$ = $1.07$, $L_y/D$ = $0.54$, and $D_c/D$ = $0.64$ which leads to an improvement of  $26.64\%$ in the value of $C_p$ over the baseline case.
From the response surfaces presented in Figure \ref{Responses}, it can be observed that the system is highly sensitive to the value of its design parameters. An increase in the cylinder diameter above its optimum value leads to a sharp decrement in $C_p$ as although a larger deflector minimizes the negative torque acting on the returning blade, such benefit is overshadowed by the flow disturbance caused by the deflector which hinders the driving torque acting on the turbine blade. Also, a larger deflector can block a part of the incoming flow striking the advancing blade which in turn further reduces the value of $C_p$. Moreover, a smaller deflector can only improve the turbine performance if it is placed suitably in the upstream direction of the returning blade but if it blocks the advancing blade then this would also lead to a diminished system performance. Similar to the effect of deflector diameter, the performance of the system diminishes when the deflector’s x and y position from the turbine is outside of the optimum range. When the deflector is placed far away from the turbine in the upstream direction it hinders its rotation by minimizing the energy of the incoming flow due to its wake and also causes flow instabilities which prevents the turbine from utilizing the maximum potential from the incoming flow. 

\begin{figure}[H]
 
\begin{tabular}{cc}
  \includegraphics[width=65mm]{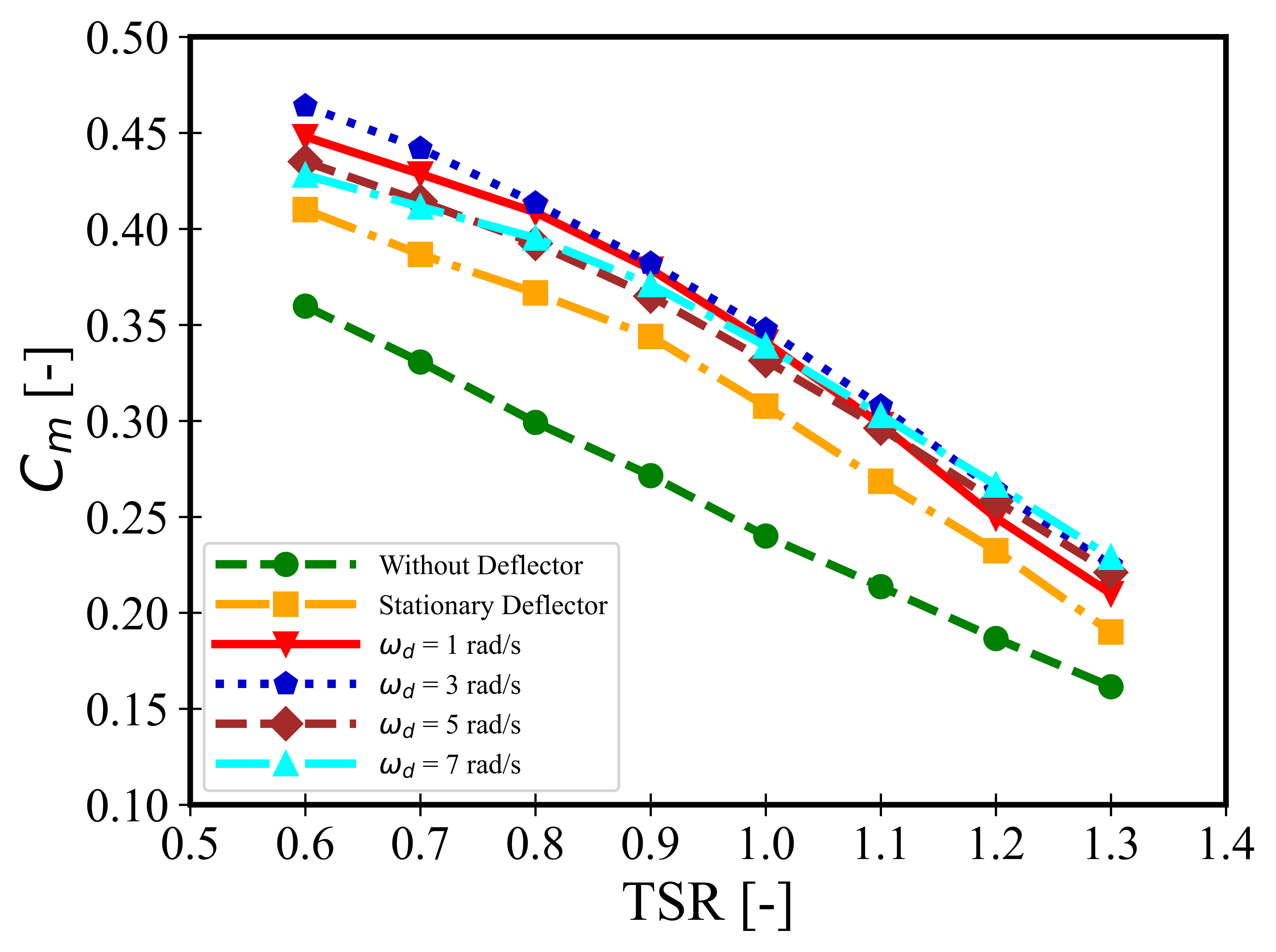} &   \includegraphics[width=65mm]{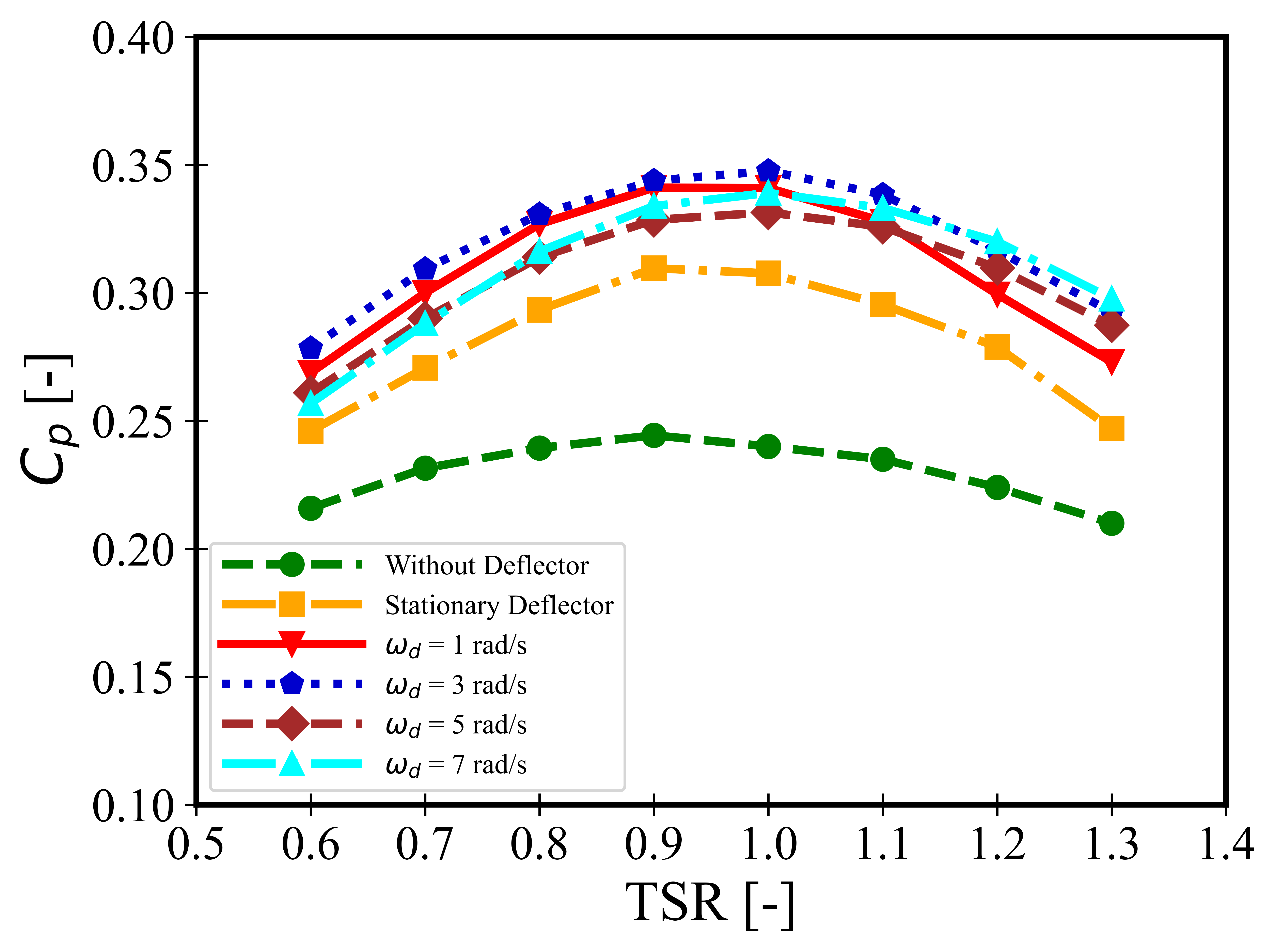} \\
(a) & (b) \\[6pt]
\end{tabular}
\caption{Variation of (a) $C_m$ and (b) $C_p$ with TSR for the without deflector case, optimized stationary deflector case, and rotating deflector case with $\omega_d$ between $1$ to $7$ $rad/s$}
\label{cm-cp-low-omega}
\end{figure}

\begin{figure}[H]
 
\begin{tabular}{cc}
  \includegraphics[width=65mm]{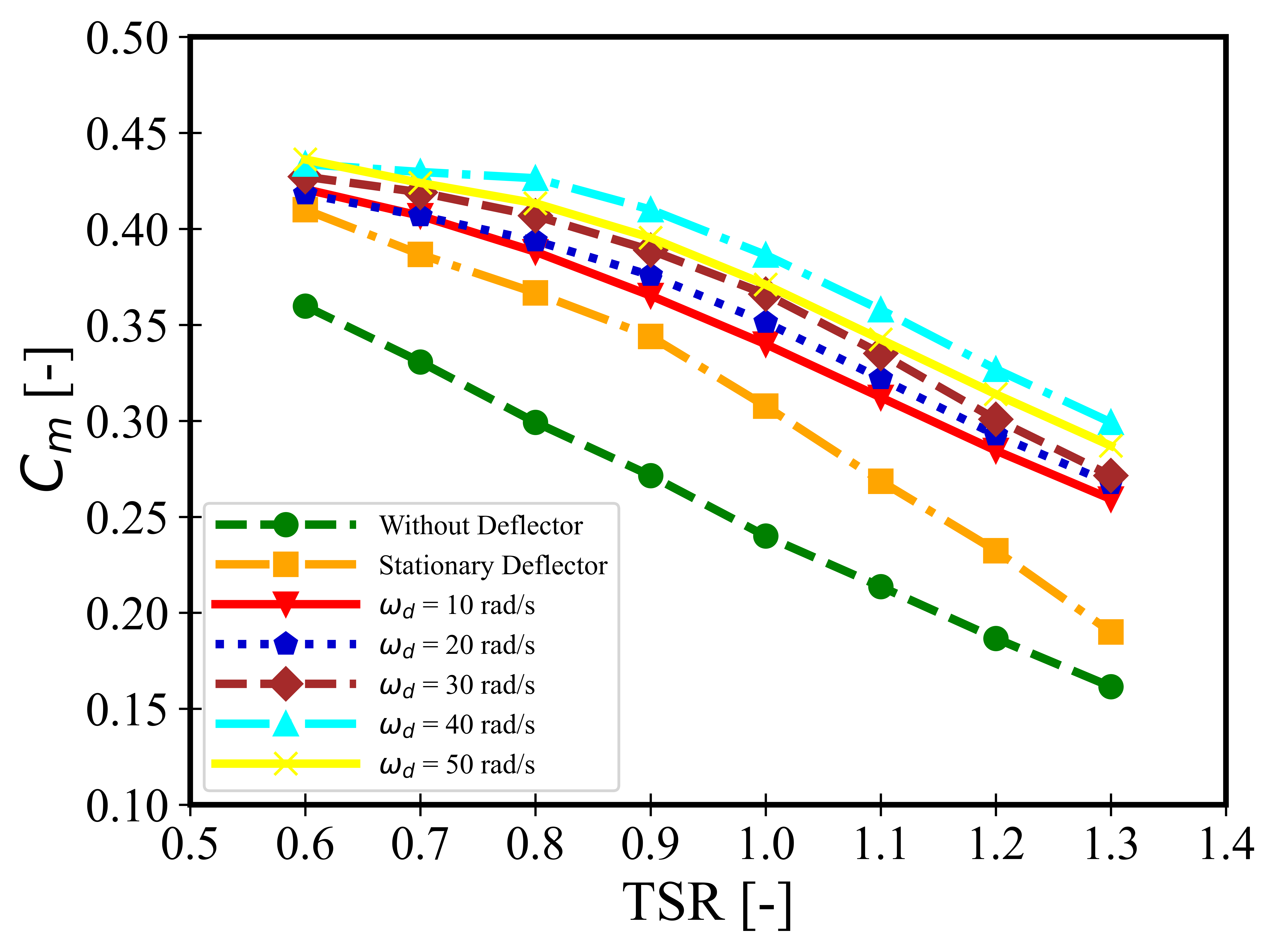} &   \includegraphics[width=65mm]{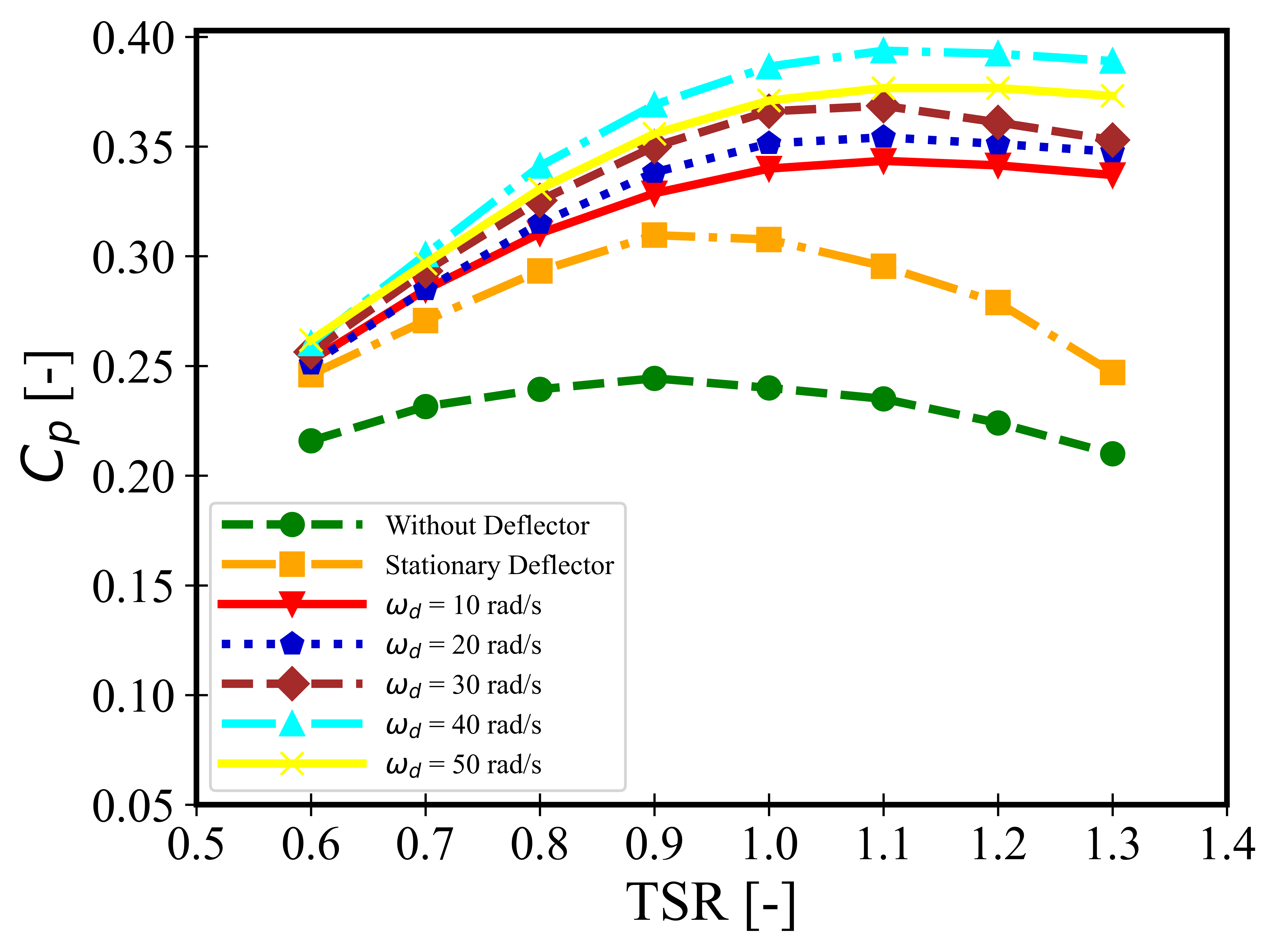} \\
(a) & (b) \\[6pt]
\end{tabular}
\caption{Variation of (a) $C_m$ and (b) $C_p$ with TSR for the without deflector case, optimized stationary deflector case, and rotating deflector case with $\omega_d$ between $10$ to $50$ $rad/s$}
\label{cm-cp-high-omega}
\end{figure}

After optimizing the geometric parameters for the turbine-deflector system, the effect of rotation of the deflector was analyzed and compared with the baseline turbine and the optimized stationary deflector-turbine system. For the present study, the angular velocity ($\omega_d$) of the deflector ranged from $1$ $rad/s$ to $50$ $rad/s$ and this range was further divided into two segments, the low and high angular velocity regions. In the low angular velocity region, $\omega_d$ was assigned the values $1$, $3$, $5$, and $7$ $rad/s$, and in the high angular velocity region, $\omega_d$ was assigned the values $10$, $20$, $30$, $40$, and $50$ $rad/s$. The sense of rotation of the deflector was taken similar to that of the turbine, i.e. in the clockwise direction. The effect of deflector rotation has been presented as a $C_m$ and $C_p$ vs TSR plot in Figures \ref{cm-cp-low-omega} and \ref{cm-cp-high-omega} for the low and high angular velocity regions. 

From Figure \ref{cm-cp-low-omega} it can be observed that all the cases with rotational velocity in the low angular velocity region provide an improvement in $C_m$ and $C_p$ at all TSR values in comparison to the stationary and the without deflector cases. Among the low rotational velocity cases, the $\omega_d$ = $3$ $rad/s$ provides the largest increment in $C_m$ and $C_p$ at all TSR values. Although the $\omega_d$ = $7$ $rad/s$ case has a lower $C_m$ and $C_p$ for the majority of the TSR range, its value becomes comparable to the $\omega_d$ = $3$ $rad/s$ case at TSR value above $1.1$. 
For the high rotational velocity region, Figure \ref{cm-cp-high-omega} shows that these cases also have a higher $C_m$ and $C_p$ at all TSR when compared with the stationary and without deflector cases. The $\omega_d$ = $40$ $rad/s$ case shows the highest improvement in $C_m$ and $C_p$ values at all TSR in comparison to other $\omega_d$ values in this region. Furthermore, at TSR values below $0.9$, the low-velocity region cases show superior performance while for TSR above $0.9$, the high-velocity cases edge over the low-velocity cases.

\begin{figure} [H]
  \centering
      \includegraphics[width=14cm]{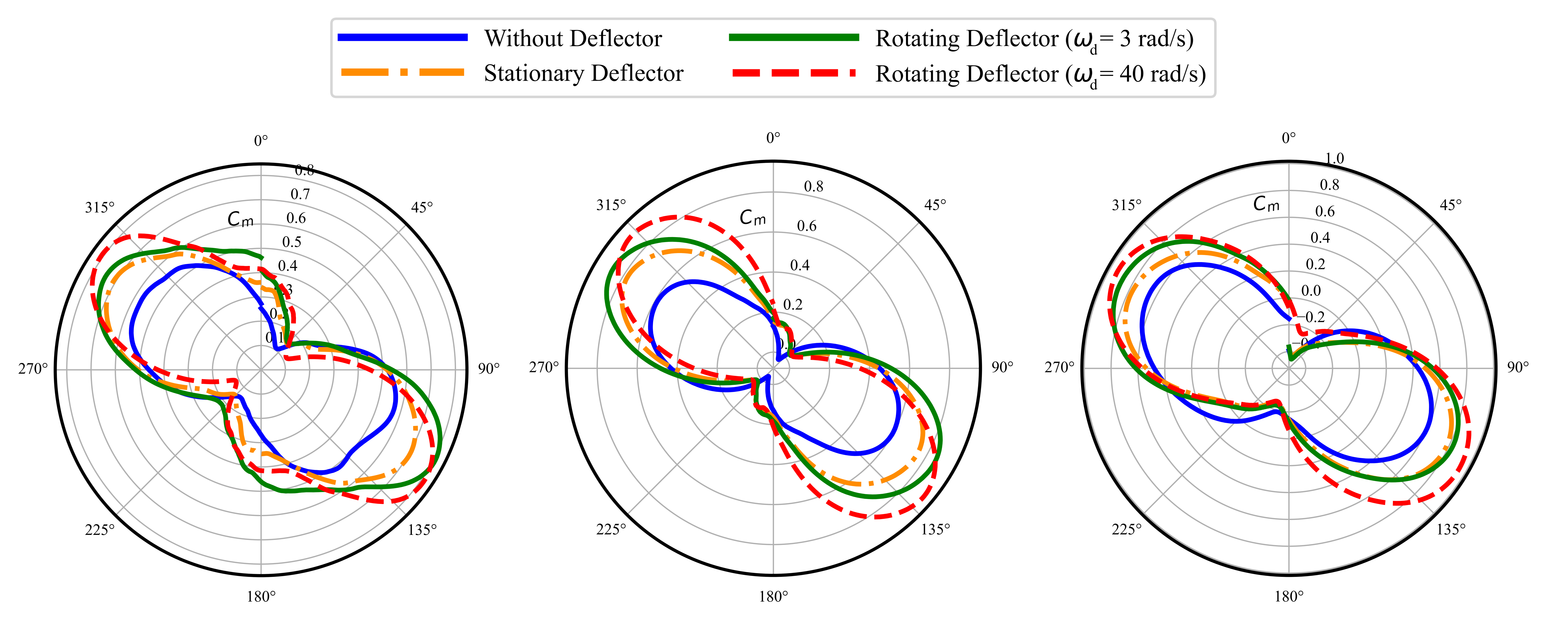}
      \caption{Polar plots representing the variation in $C_m$ over one rotational cycle of the turbine at TSR values of (a) $0.6$, (b) $0.9$, and (c) $1.3$ }
      \label{radial plot}
\end{figure}

The variation in the instantaneous value of $C_m$ for the without deflector, stationary deflector, $\omega_d$ = $3$ $rad/s$, and $\omega_d$ = $40$ $rad/s$ cases at TSR values of $0.6$, $0.9$, and $1.3$ are presented in the form of polar plots in Figure \ref{radial plot}. A phase shift in the $C_m$ behavior can be observed at different TSR values and also there is a slight phase shift in $C_m$ among the four turbine-deflector configurations for the same TSR values.  For all three TSR values, the $\omega_d$ = $40$ $rad/s$ has a higher peak $C_m$ value followed by the $\omega_d$ = $3$ $rad/s$ case, then by the stationary deflector case, and the least value is for the without deflector case. Also, the peak value of $C_m$ is reached when the turbine is at approximately $120^{\circ}$ Azimuth angle. 

At TSR $0.9$, the without deflector case faces negative or net retarding torque for a small Azimuth angle range while such effect is mitigated with the use of stationary as well as rotating deflectors. Furthermore, for TSR $1.3$, only the $\omega_d$ = $40$ $rad/s$ case faces a positive torque throughout the rotational cycle of the turbine. For TSR $0.6$ and TSR $1.3$, the torque profile at the peak values of $C_m$ is broader for the without deflector case than it is for the stationary and rotating deflector cases, while this effect is opposite in the case of TSR $0.9$.

At TSR $0.6$, the $\omega_d$ = $3$ $rad/s$ case and the without deflector case have a different $C_m$ value at the end of one rotational cycle than when it was at the start of that cycle. Similarly, for TSR $1.3$, all four cases have different start and end $C_m$ values. This shows that the frequency of torque variation changes with TSR and turbine-deflector configuration.

\subsection{Flow Investigation}
\begin{figure} [H]
  \centering
      \includegraphics[width=14cm]{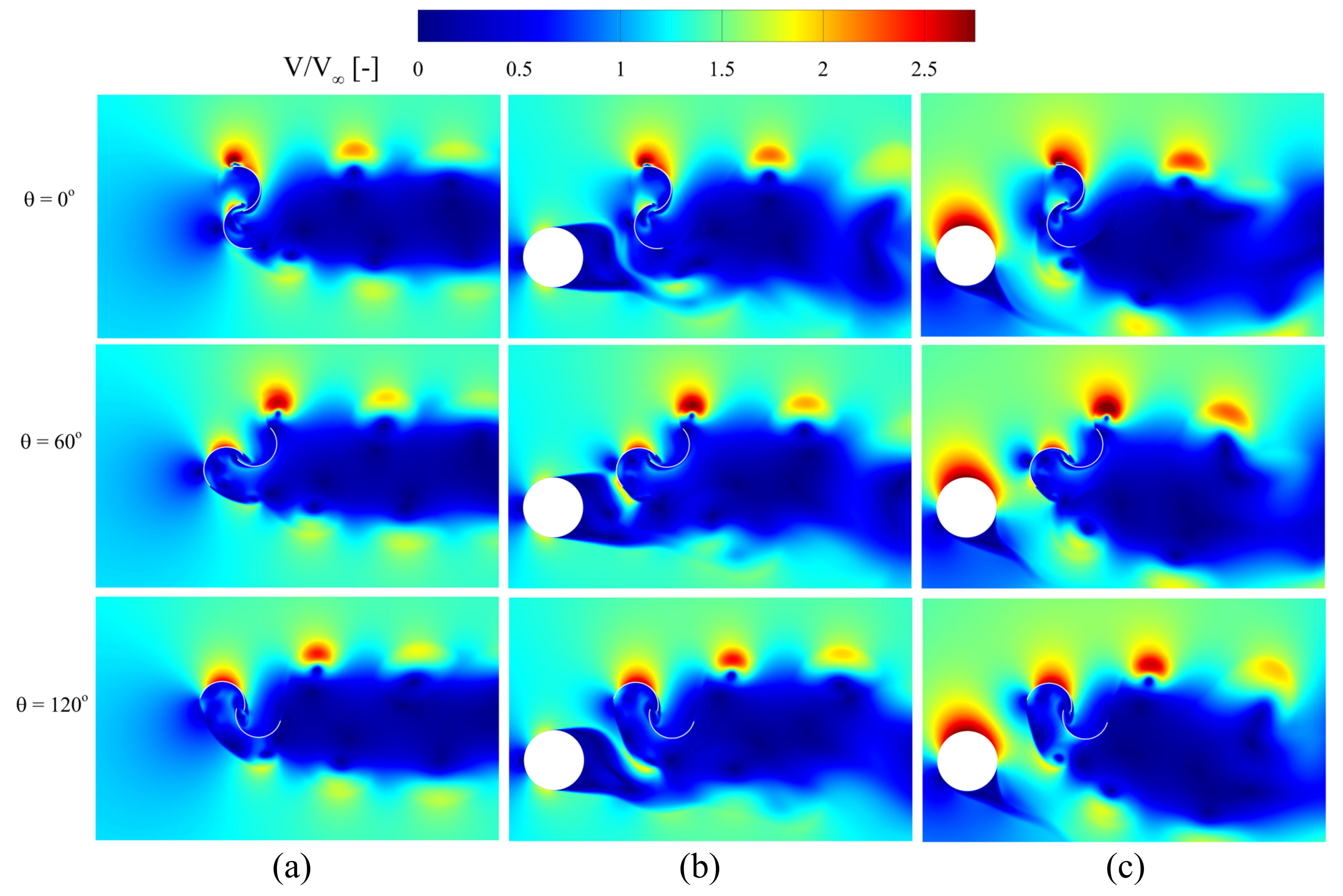}
      \caption{Normalized velocity distribution at different Azimuth angles for (a) without deflector, (b) stationary deflector, and (c) rotating deflector ($\omega_d$ = $40$ $rad/s$) cases }
      \label{velocity plot}
\end{figure}

To understand the flow dynamics responsible for improvements in turbine performance due to the use of stationary and rotating upstream deflectors, a detailed flow analysis is presented. At first, the distribution of normalized velocity for the without deflector, stationary deflector, and rotating deflector ($\omega_d$ = $40$ $rad/s$) cases at different Azimuth angles ($\theta$) is presented in Figure \ref{velocity plot}. In all three cases at $\theta$ = $0^o$, a high-velocity region can be observed at the tip of the advancing blade. This is due to the combined effect of the incoming flow and the clockwise rotation of the turbine. Furthermore, when the turbine is at $\theta$ = $60^o$, a high-velocity region develops near the mid of the retarding blade at the windward side. The incoming flow gets accelerated due to the convexity of the blade and also the clockwise rotation of the turbine adds to this effect. Similarly, at $\theta$ = $120^o$, a high-velocity zone develops near the middle of the advancing blade at the leeward side. 
Due to the presence of a deflector in Figure \ref{velocity plot}(b) and Figure \ref{velocity plot}(c), an obstruction to the flow striking the retarding blade is created which reduces the impact of the retarding torque. Furthermore, in the wake of the rotating deflector, the flow is deflected even further downwards away from the turbine’s retarding blade which explains its maximum performance.

\begin{figure} [H]
  \centering
      \includegraphics[width=14cm]{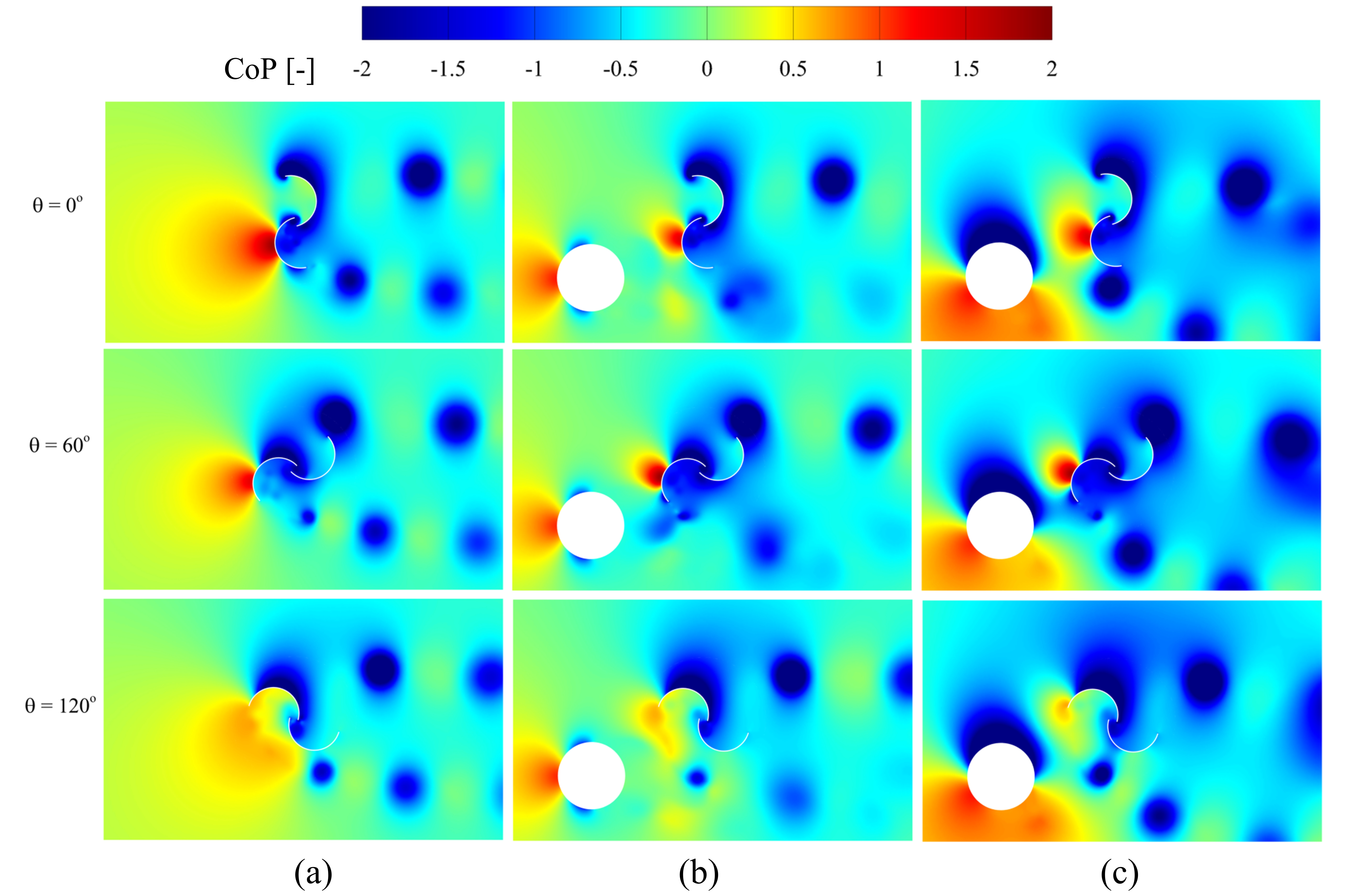}
      \caption{Pressure Coefficient (CoP) distribution at different Azimuth angles for (a) without deflector, (b) stationary deflector, and (c) rotating deflector ($\omega_d$ = $40$ $rad/s$) cases}
      \label{pressure plot}
\end{figure}

To understand the pressure variation around the system, the distribution of pressure coefficient ($CoP$ = $P/0.5^*\rho^*V_{\infty}^2$) around the without deflector, stationary deflector, and rotating deflector ($\omega_d$ = $40$ $rad/s$) cases at different Azimuth angles is presented in Figure \ref{pressure plot}. For the advancing blade of the turbine, a high-pressure region around the blade portion facing the windward side and a low-pressure region at the leeward side can be observed. This is caused by the sharp curvature change along the blade profile due to which there is an airflow separation and a loss in pressure which contributes to the pressure or form drag acting on the advancing blade, aiding it to drive in the clockwise direction. A similar effect occurs for the retarding blade, but the moment caused by the acting forces tend to drive it in the anticlockwise direction and thus oppose the rotation of the turbine. 
Due to the use of stationary and rotating deflectors, it can be clearly observed from Figure \ref{pressure plot} that the high-pressure region around the windward side of the retarding blade is significantly smaller than it is for the without deflector case at all Azimuth angles. This in turn contributes to the higher $C_m$ and $C_p$ with the use of deflectors. Furthermore, on comparing the stationary and rotating deflector case, a suction pressure region can also be observed alongside the high-pressure region at the windward side of the retarding blade with the use of rotating deflectors. This region develops due to the clockwise rotation of the deflector which increases the velocity of the air at the topmost point of the deflector which in turn leads to a decrease in the air pressure. Due to this suction region, the effect of the high-pressure region near retarding blade is further reduced and ultimately contributing towards an even higher power output using rotating deflectors.

\begin{figure} [H]
  \centering
      \includegraphics[width=14cm]{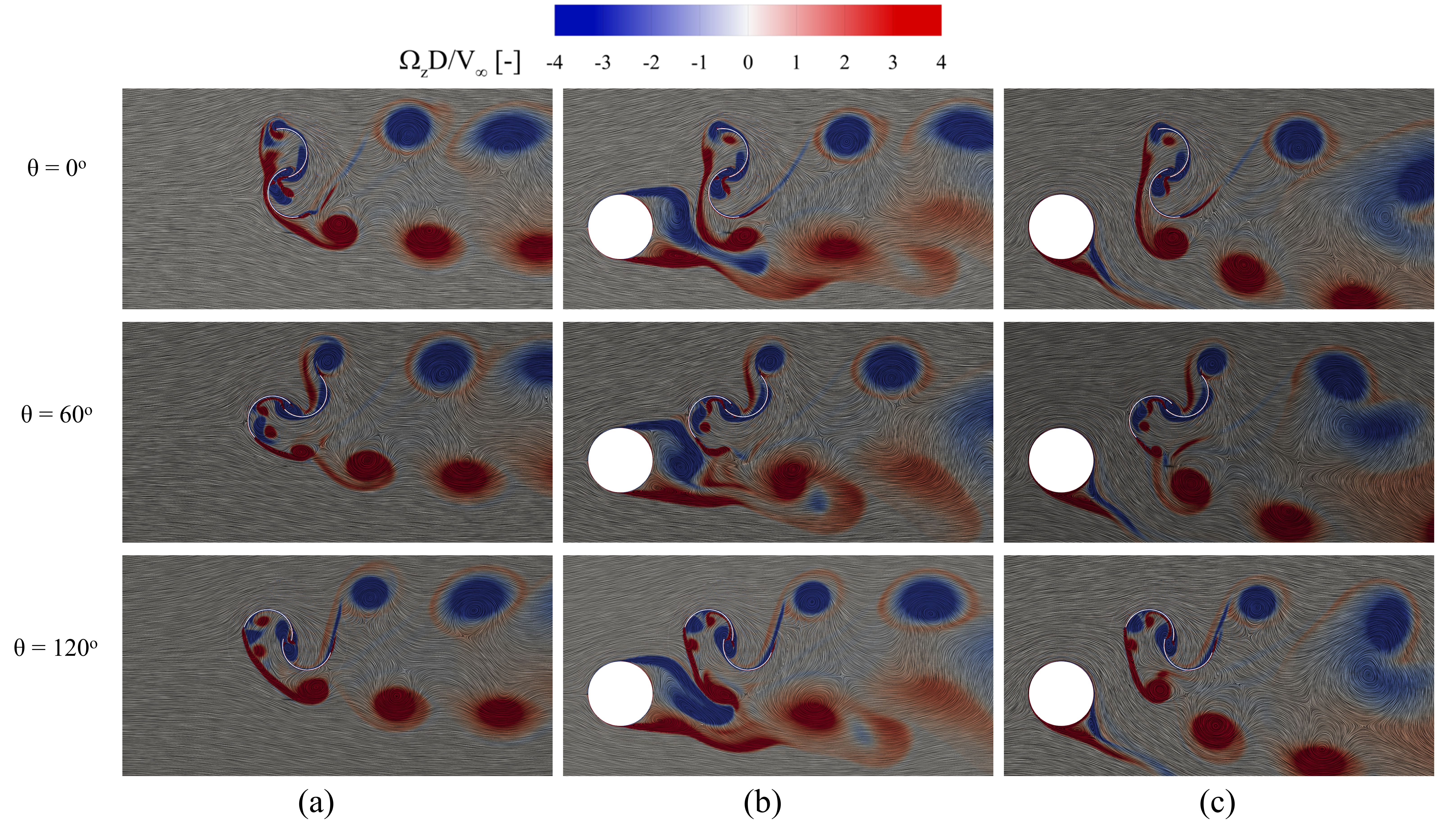}
      \caption{Line-Integral Convolution plots illustrating the normalized Z-Vorticity distribution at different Azimuth angles for (a) without deflector, (b) stationary deflector, and (c) rotating deflector ($\omega_d$ = $40$ $rad/s$) cases}
      \label{LIC plot}
\end{figure}

Next, to get clarity over the vortex interactions and flow dynamics, normalized Z-Vorticity distribution is presented on Line-Integral Convolution (LIC) plots for the without deflector, stationary deflector, and rotating deflector ($\omega$ = $40$ $rad/s$) cases at different Azimuth angles in Figure \ref{LIC plot}. These LIC plots can be interpreted as space-filling streamline plots. For the without deflector case, when the turbine rotates from $0^o$ to $60^o$, a clockwise vortex is formed at the tip of the advancing blade which gets shed when the turbine rotates to $120^o$ Azimuth angle. Similarly, at the tip of the retarding blade, an anticlockwise vortex is formed and then it gets shed due to the rotation of the blades. After getting shed from the turbine, these vortices get convected downstream.  

For the stationary deflector case, a pair of counter-rotating vortices are formed behind the deflector and its shedding into a Karman vortex street is prevented due to the presence of the Savonius turbine in the downstream direction. Furthermore, the vortices in the deflector wake interact with the anti-clockwise vortex generated by the retarding blade. The clockwise vortex from the upper surface of the deflector impinges with the anti-clockwise vortex from the retarding blade and forms a vortex pair which gets convected downstream. 
In comparison to the stationary deflector case, the interaction between the vortices generated by the deflector and the turbine is comparatively less. This is due to the high rotational velocity of the deflector which shifts the Kutta condition near the lower end of the deflector facing the leeward side. Moreover, the rotational velocity of the deflector also deflects the vortices from the retarding blade in the downward direction. As evident in Figure \ref{LIC plot}(c), the size of the deflector wake and the recirculation zone is also significantly reduced. This reduction in wake size of the deflector, shift in its Kutta condition downwards, and the deflection of incoming flow towards the retarding blade on the downward side is majorly responsible for the high power output of the rotating deflector case in comparison to the other two cases.

\subsection{Energy Consumption}
\begin{figure} [H]
 
\begin{tabular}{cc}
  \includegraphics[width=65mm]{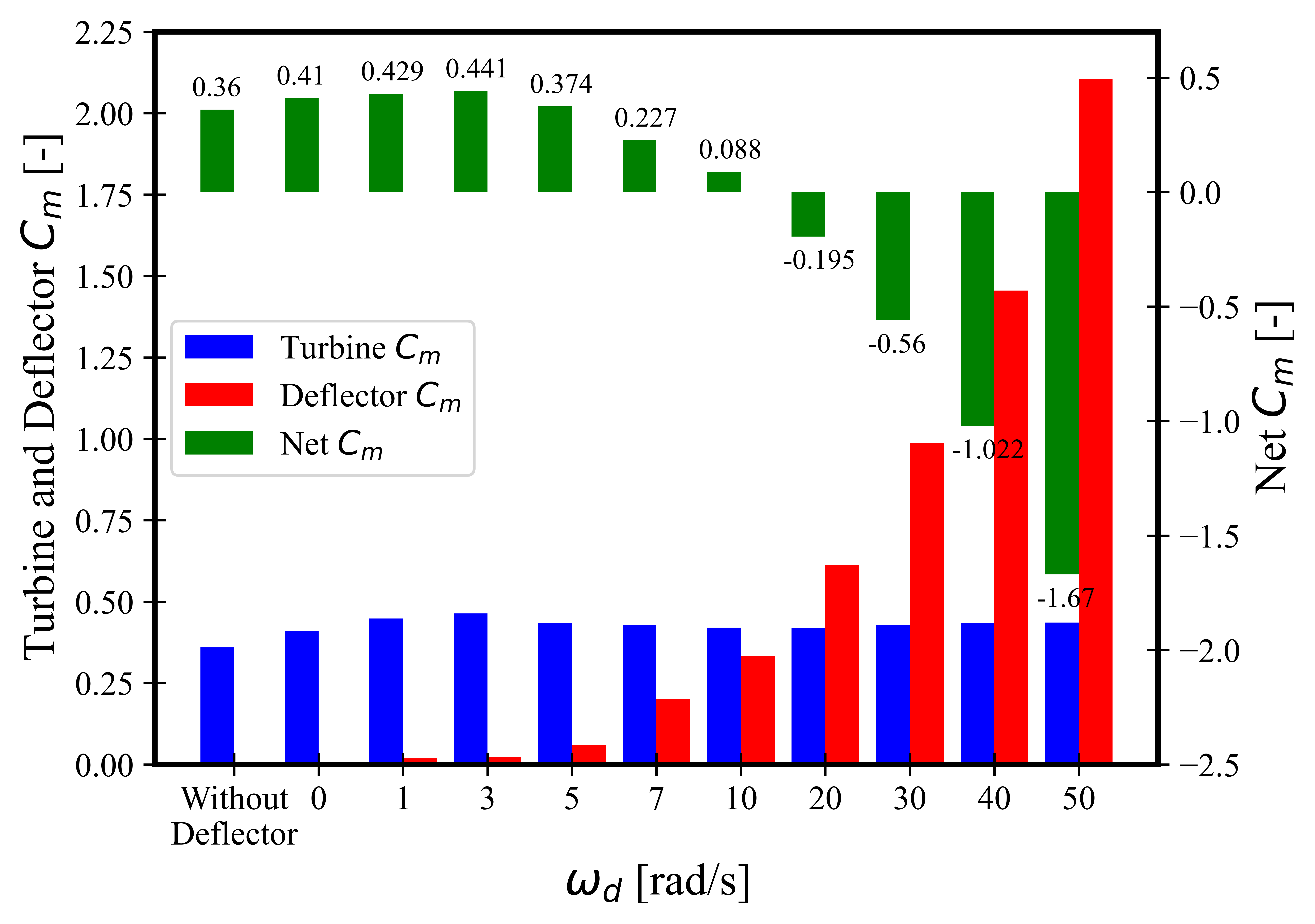} &   \includegraphics[width=65mm]{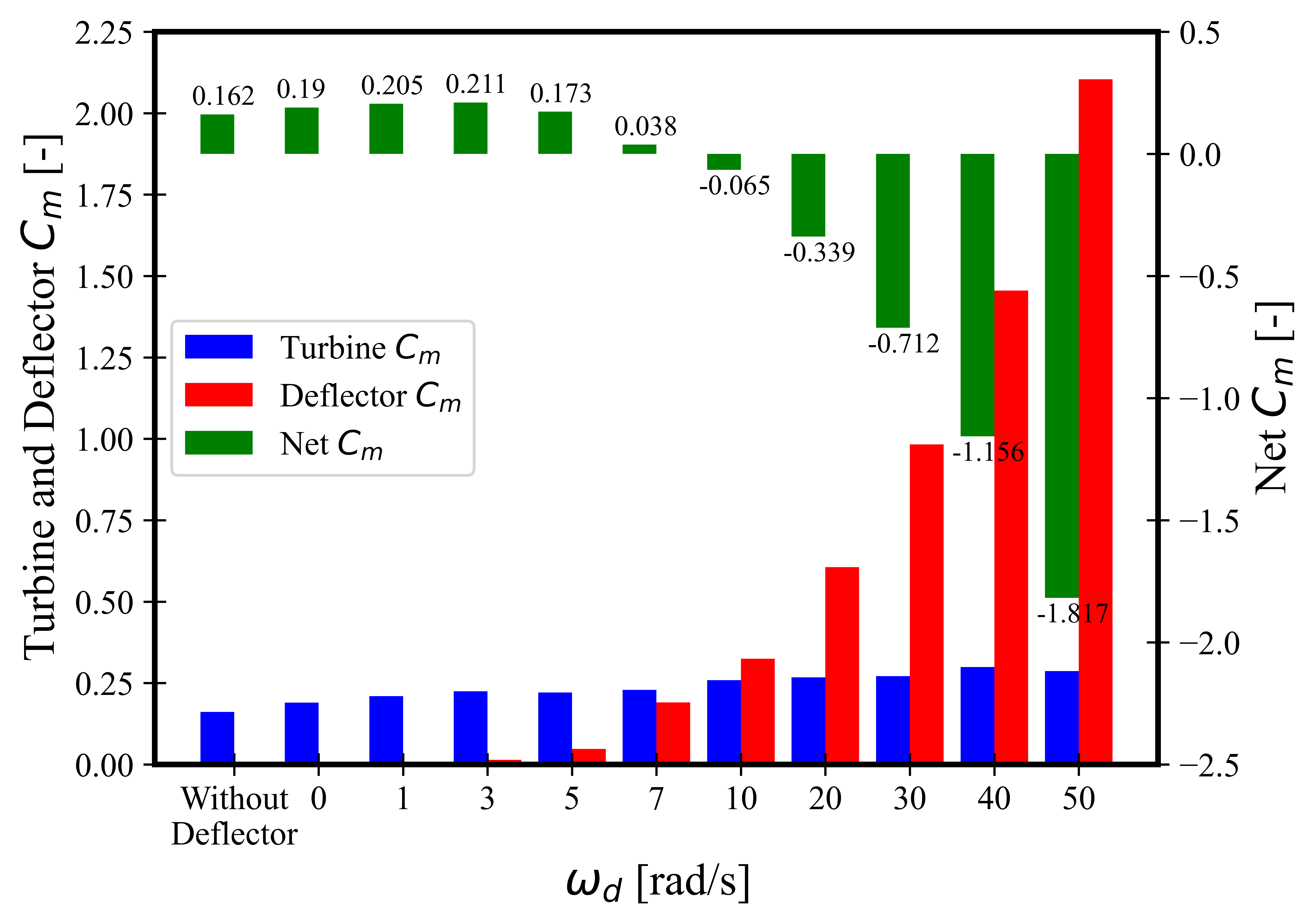} \\
(a) & (b) \\[6pt]
\end{tabular}
\caption{Net $C_m$ output at TSR (a) 0.6 and (b) 1.3}
\label{energy-consumption}
\end{figure}

All the results presented until now have shown the benefit of using rotating deflectors over stationary deflectors. But it must be noted that for rotating the deflectors, external work is required which must be taken into account while calculating the net torque output from the turbine-deflector system. To quantify this, the net $C_m$ values at different TSRs have been presented in Figure \ref{energy-consumption}.  
At TSR $0.6$, it can be observed from Figure \ref{energy-consumption}(a) that only for $\omega_d$ $<$ $20$ $rad/s$, the system shows a positive torque output, and for the rest of the cases, external work must be done to drive the system. Furthermore, at $\omega_d$ = $3$ $rad/s$, the rotating deflector-turbine system shows the maximum net $C_m$ output which is $7.56\%$ higher than the stationary deflector case and $22.5\%$ higher than the without deflector case.
At TSR $1.3$, Figure \ref{energy-consumption}(b) shows that positive torque output is achieved when $\omega_d$ $<$ $10$ $rad/s$. Similar to the TSR $0.6$ case, peak $C_m$ value is achieved for $\omega_d$ = $3$ $rad/s$ rotating deflector case and its value is $9.95\%$ higher than the stationary deflector case and $30.25\%$ higher than the without deflector case.

\section{Conclusion}
Through this study, a robust and efficient design optimization framework has been proposed to improve the performance of a Savonius wind turbine equipped with a cylindrical deflector. A multi-faceted approach involving CFD simulations of the turbine-deflector configuration, surrogate modeling, and optimization using Quantum-inspired metaheuristic algorithms was used. Parameters considered for the optimization study were the horizontal and vertical distances of the center of the cylindrical deflector from the turbine shaft and the diameter of the deflector. After parameterizing the turbine-deflector system, sixty distinct design points were generated through Latin Hypercube Sampling, and data for training the surrogate model was gathered using URANS simulations. Subsequently, six different surrogate models were evaluated based on their accuracy in approximating the results of CFD simulations and the Kriging surrogate model was found to be the best representation of the system. Next, on the response surface generated by the Kriging model, the Quantum-based Salp Swarm Algorithm (QSSO) was used to compute the system parameters corresponding to the global maxima of the objective function, i.e. $C_p$. Additionally, the performance of nine other state-of-the-art metaheuristic algorithms was extensively compared with the QSSO algorithm. The QSSO algorithm was observed to converge much faster than all other investigated algorithms, while also escaping the local maxima and reaching the global maxima. \par

Upon employing the QSSO algorithm, it was found that design parameters having values, $L_x/D$ = $1.07$, $L_y/D$ = $0.54$, and $D_c/D$ = $0.64$ correspond to the optimal configuration for the stationary deflector-turbine system. Using this configuration, the value $C_p$ improved from $0.244$ to $0.309$, an improvement of $26.64\%$ in comparison to the baseline turbine without any deflector at a TSR of $0.9$. After optimizing the geometric parameters of the system and addressing the shortcomings of the stationary deflector, the effect of deflector rotation on the system performance was evaluated. Deflectors with angular velocity ($\omega_d$) ranging from $1$ to $50$ $rad/s$ were studied. It was found that for $\omega_d$ = $3$ $rad/s$ and $40$ $rad/s$, the system gave improvements in $C_p$ over approximately the entire TSR range evaluated in comparison to the without deflector and the optimized stationary deflector case. Using a deflector with $\omega_d$ = $3$ $rad/s$ gave a $Cp$ value of $0.344$, an improvement of $40.98\%$ and $11.33\%$ compared to the without deflector and optimized stationary deflector case. Similarly, using a deflector with $\omega_d$ = $40$ $rad/s$ gave a $Cp$ value of $0.369$, an improvement of $51.23\%$ and $19.42\%$ compared to the without deflector and optimized stationary deflector case. Apart from computing the increment in power and torque output of the turbine using rotating deflectors, the net torque output of the overall system was also calculated. Based on the results, the $\omega_d$ = $3$ $rad/s$ case showed maximum net torque output at TSR $0.6$ and $1.3$, while the high angular velocity cases showed negative net torque output at both the evaluated TSRs. \par

Besides its primary utilization in optimizing design parameters for the turbine-deflector system, the framework introduced demonstrates a promising potential for wider application in optimizing diverse aerodynamic components utilized across wind energy, aerospace, and automotive sectors. Given the foundational principles and methodologies integrated into the framework, it is reasonable to infer that its relevance extends beyond the specific focus of this study.

\section*{CRediT authorship contribution statement}
\textbf{Paras Singh:} Conceptualization, Methodology, Software, Validation, Formal analysis, Data Curation, Writing - Original Draft, Visualization. \textbf{Vishal Jaiswal:} Conceptualization, Methodology, Software, Validation, Formal analysis, Data Curation, Writing - Original Draft, Visualization. \textbf{Subhrajit Roy:} Methodology, Validation, Formal analysis, Data Curation, Writing - Original Draft, Visualization. \textbf{Aryan Tyagi:} Conceptualization, Software, Formal analysis, Writing - Original Draft. \textbf{Gaurav Kumar:} Writing - Review \& Editing, Supervision. \textbf{Raj Kumar Singh:} Writing - Review \& Editing, Supervision, Project administration.

\section*{Declaration of competing interest}
The authors declare that they have no conflict of interest and no funding was received for conducting this study.

\section*{Data availability}
The source code developed for this work, the setup, and the scripts required to run the optimizations are available at \\\emph{\href{https://github.com/VizArcher/Quantum-Based-Metaheuristics}{\textcolor{blue}{https://github.com/VizArcher/Quantum-Based-Metaheuristics}}}




\bibliographystyle{elsarticle-num} 
\bibliography{refs_bibtex}





\end{document}